\documentclass[]{aastex631}

\usepackage{amsmath}
\usepackage{makecell}

\usepackage{scalerel}
\newcommand\HII{H{\sc ii}}
\newcommand\HI{H{\sc i}}
\newcommand\Halpha{H{$\alpha$}}
\newcommand{\RN}[1]{\textup{\uppercase\expandafter{\romannumeral#1}}}

\def\NHUNIT{\ifmmode {\rm \,cm^{-2}} \else $\rm \,cm^{-2}$ \fi}
\newcommand{\bperp}{$B_{\textsc{POS}}$} 

\received{}
\revised{}
\accepted{}
\submitjournal{ApJ}

\shorttitle{NGC~6334 bubbles}
\shortauthors{Tahani et al.}
\graphicspath{{./}{figures/}}

\begin{document}

\title{JCMT BISTRO Observations: Magnetic Field Morphology of Bubbles Associated with NGC 6334} 

\correspondingauthor{Mehrnoosh~Tahani}
\email{mtahani@stanford.edu}

\author[0000-0001-8749-1436]{Mehrnoosh Tahani}
\affiliation{Banting and KIPAC Fellowships: Kavli Institute for Particle Astrophysics \& Cosmology (KIPAC), Stanford University, Stanford, CA 94305, USA}
\affiliation{Dominion Radio Astrophysical Observatory, Herzberg Astronomy and Astrophysics Research Centre, National Research Council Canada, P. O. Box 248, Penticton, BC V2A 6J9 Canada}

\author[0000-0002-0794-3859]{Pierre Bastien}
\affiliation{Institut de Recherche sur les Exoplan\`etes (iREx), Universit\'e de Montr\'eal, D\'epartement de Physique, 1375, Avenue Th\'er\`ese-Lavoie-Roux, Montr\'eal, QC, H2V 0B3, Canada}
\affiliation{Centre de Recherche en Astrophysique du Qu\'ebec (CRAQ), Universit\'e de Montr\'eal, D\'epartement de Physique, 1375, Avenue Th\'er\`ese-Lavoie-Roux, Montr\'eal, QC, H2V 0B3, Canada}

\author[0000-0003-0646-8782]{Ray S. Furuya}
\affiliation{Institute of Liberal Arts and Sciences, Tokushima University, Minami Jousanajima-machi 1-1, Tokushima 770-8502, Japan}

\author[0000-0002-8557-3582]{Kate Pattle}
\affiliation{Department of Physics and Astronomy, University College London, WC1E 6BT London, UK}

\author[0000-0002-6773-459X]{Doug Johnstone}
\affiliation{Department of Physics and Astronomy, University of Victoria, Victoria, BC V8P 1A1, Canada}
\affiliation{Herzberg Astronomy and Astrophysics Research Centre, National Research Council of Canada, 5071 West Saanich Rd, Victoria, BC V9E 2E7, Canada}

\author{Doris Arzoumanian}
\affiliation{National Astronomical Observatory of Japan, National Institutes of Natural Sciences, Osawa, Mitaka, Tokyo 181-8588, Japan}

\author[0000-0001-8746-6548]{Yasuo~Doi}
\affiliation{Department of Earth Science and Astronomy, Graduate School of Arts and Sciences, The University of Tokyo, 3-8-1 Komaba, Meguro, Tokyo 153-8902, Japan}

\author[0000-0003-1853-0184]{Tetsuo Hasegawa}
\affiliation{National Astronomical Observatory of Japan, National Institutes of Natural Sciences, Osawa, Mitaka, Tokyo 181-8588, Japan}

\author[0000-0003-4366-6518]{Shu-ichiro Inutsuka}
\affiliation{Department of Physics, Graduate School of Science, Nagoya University, Furo-cho, Chikusa-ku, Nagoya 464-8602, Japan}

\author[0000-0002-0859-0805]{Simon Coud\'e}
\affiliation{SOFIA Science Center, Universities Space Research Association, NASA Ames Research Center, M.S. N232-12, Moffett Field, CA 94035, USA}
\affiliation{Centre de Recherche en Astrophysique du Qu\'ebec (CRAQ), Universit\'e de Montr\'eal, D\'epartement de Physique, C.P. 6128 Succ. Centre-ville, Montr\'eal, QC H3C 3J7, Canada}

% ...

\begin{abstract}
We study the \HII\ regions associated with the NGC~6334  molecular cloud observed in the sub-millimeter and taken as part of the B-fields In STar-forming Region Observations (BISTRO) Survey. In particular, we investigate the polarization patterns and magnetic field morphologies associated with these \HII\ regions. 
Through polarization pattern and pressure calculation analyses, several of these bubbles indicate that the gas and magnetic field lines have been pushed away from the bubble, toward an almost tangential (to the bubble) magnetic field morphology. In the densest part of NGC~6334, where the magnetic field morphology is similar to an hourglass, the polarization observations do not exhibit observable impact from \HII\ regions. We detect two nested radial polarization patterns in a bubble to the south of NGC~6334 that correspond to the previously observed bipolar structure in this bubble. Finally, using the results of this study, we present steps (incorporating computer vision; circular Hough Transform) that can be used in future studies to identify bubbles that have physically impacted magnetic field lines.
\end{abstract}


\section{Introduction} 
\label{sec:intro}

Ionized atomic hydrogen (\HII ) regions around massive stars are evidence of recent star formation activity. 
Expansion of these \HII\ regions and their interactions with molecular clouds~\citep{Tremblinetal2012a} can also lead to increased density~\citep{Thomsonetal2012, Inutsukaetal2015} along their edges and the initiation of a new star-formation sequence~\citep[e.g.,][]{Elmegreen1998, Deharvengetal2005, Zavagnoetal2010b, Zavagnoetal2010, Chenetal2022}. Conversely, stellar outflows combined with magnetic fields can reduce the efficiency of massive star formation significantly~\citep{NakamuraLi2007, Wangetal2010, Kochetal2012a, Kochetal2012b, Federrath2015, KrumholzFederrath2019, Chenetal2022}. In magnetohydrodynamic simulations, for instance, magnetic fields enable outflows to travel farther than distances in hydrodynamic simulations, resulting in turbulent motions and decreased star-formation efficiency at greater distances~\citep[e.g.,][]{OffnerLiu2018, KrumholzFederrath2019}. The initial morphology of the magnetic field in the cloud relative to the \HII\ regions also influences the star-formation efficiency~\citep{Chenetal2022}. Before examining the star-formation activities caused by \HII\ regions, it is necessary to study the magnetic field morphology of these regions and the influence of \HII\ regions on the magnetic fields of the parent molecular cloud.

The physical sizes of \HII\ regions range from ultra-compact ($<0.1$\,pc) to  evolved~\citep[$>1$\,pc;][]{Russeiletal2016}, enabling us to estimate their age, ionizing flux, and density~\citep[e.g.,][]{Osterbrocketal2006, Andersonetal2014}. These regions can be identified in mid-infrared (MIR) observations as bubble- or shell-like structures, which are often asymmetrical in shape~\citep{Comeron1997} and are formed as a result of winds and radiation from the massive central stars. The stellar winds and radiation result in expanding \HII\ regions, which sweep material on their shells and consequently alter the magnetic field morphology of their  environment (due to flux freezing) compared to the rest of the parent cloud. 

While \HII\ region age estimates are required for studying star formation, their ages can be underestimated if magnetic fields are ignored~\citep{Chenetal2022}. This underestimation of age is due to the fact that strong magnetic fields resist the \HII\ expansion perpendicular to the field direction~\citep[which may also result in ovoid-shaped bubbles;][]{Chenetal2022}. The relative orientation of magnetic field with the expansion of the \HII\ region influences not only the density of the shell but also its fragmentation and core formation~\citep[][]{Chenetal2022}. Some studies have examined the role and significance of magnetic fields in cloud evolution and star formation and have shown that the magnetic fields can be aligned tangential to the shells of  evolved \HII\ regions~\citep[e.g.,][]{PereyraMagalhaes2007, Tangetal2009, Chenetal2012, Santosetal2014, Fisseletal2016, Soametal2017, Chenetal2017, Eswaraiahetal2017, Soametal2018, Pattleetal2018, Dewangan2018, Eswaraiahetal2020,  Konyvesetal2021, Devarajetal2021, Lopezetal2021, HoangetalThu2022, Chungetal2022}. This tangential field morphology is also predicted in some theoretical simulations~\citep{Krumholzetal2007, Henneyetal2009, Arthuretal2011, Ntormousietal2017}. However, the complete morphology and role of magnetic fields in relation to \HII\ regions are not yet well understood. 

If \HII\ regions orient and order the field lines tangential to them, this will also appear as tangential magnetic fields when projected onto the plane of the sky and will be accompanied by higher polarization fractions. These ordered field lines will experience less deviation from tangential morphology, leading to less depolarization and a higher polarization fraction. 
To better understand the morphology of magnetic fields associated with \HII\ regions, here we study the magnetic fields in the NGC~6334 molecular cloud and its \HII\ regions. In this study, we investigate the presence or absence of tangential magnetic field morphology in \HII\ regions, as well as whether the presence of such fields is accompanied by an increase in polarization fraction. Determining the presence of this morphology may reveal the significance of \HII\ regions in the evolution of cloud's magnetic fields. Furthermore, by understanding how \HII\ regions may influence the morphology of magnetic fields in their parental cloud, we may also be able to determine the initial magnetic field morphology of the parental cloud prior to its evolution. Knowing this initial magnetic field morphology allows us to compare it to the magnetic field morphology predicted by cloud-formation scenarios. 

Numerous \HII\ regions are associated with NGC~6334, which is a massive star-forming molecular cloud, on the inner edge of the Sagittarius-Carina arm~\citep{Russeiletal2012}, with core masses ranging from 200 to 2000M$_{\odot}$~\citep[][]{Zernickeletal2013} and a total mass of a few 10$^5$M$_{\odot}$~\citep[][]{Andreetal2016}. These \HII\ regions have influenced the gas dynamics in NGC~6334~\citep{Schneider2020} and some of them are well-known and optically observed~\citep[e.g.,][]{Gum1955, PersiTapia2008}. The presence of numerous \HII\ regions along the filamentary NGC~6334 molecular cloud, which were initially detected by ~\citet[][]{Rodriguezetal1982} and have been extensively studied, is not unique to this region and has been similarly observed in other molecular clouds~\citep[e.g., NGC~6357;][]{Russeiletal2010}. \citet[][]{Rodriguezetal1982} estimate the luminosity of the \HII\ regions within NGC~6334 and identify the Zero Age Main Sequence stellar type required to generate the estimated luminosity (see their Table~2). Subsequent studies identify a number (or cluster) of stars that may have contributed to the formation of the NGC~6334 \HII\ regions~\citep[e.g.,][]{Tapiaetal1996}.

NGC~6334 and its sub-regions have been observed with various radio, millimeter, and sub-millimeter facilities, including the Atacama Large Millimeter/sub-millimeter Array~\citep[ALMA; e.g.,][]{Sadaghianietal2020, Cortesetal2021}, the Very Large Array~\citep[VLA; e.g.,][]{Rodriguezetal2014}, and the Sub-Millimeter Array~\citep[SMA; e.g.,][]{Zhangetal2014}.  NGC~6334 is actively forming stars and includes a $\sim 10$\,pc filament~\citep[main ridge;][]{Loughranetal1986, Russeiletal2010, Shimajirietal2019} with an average density significantly higher than most filaments observed in the Gould Belt~\citep{Andreetal2016}. \citet{Wuetal2014} determined a parallax distance of $1.35^{+0.15}_{-0.13}$\,kpc for NGC~6334, $\sim 20\%$ smaller than its near kinematic distance of 1.7\,kpc~\citep{Wuetal2014} and photometric distance of $1.74 \pm 0.31$\,kpc~\citep{Neckel1978}.

To investigate the magnetic field morphology of NGC~6334 and its \HII\ regions, we observe polarized dust emission at 850\,$\mu$m, using the James Clerk Maxwell Telescope\footnote{\url{https://www.eaobservatory.org/jcmt/}} (JCMT) near the summit of Mauna Kea.  
Probing the plane-of-sky magnetic field (\bperp ) morphology using polarized dust emission is possible due to the alignment of the short axis of amorphous dust grains~\citep[e.g.,][]{Draine2009} with the magnetic field, which has been observed at various wavelengths~\citep[e.g.,][]{PattleFissel2019}. This alignment is explained through radiative torques~\citep[RAT;][]{Draineetal1997, Lazarian2007, LazarianHoang2007, Anderssonetal2015, HoangLazarian2016}. 

Previously, \citet[][hereafter Paper~\RN{1}]{Arzoumanian2020} studied the magnetic fields within NGC~6334 and the region's filamentary nature. We focus on its \HII\ regions in this study. We discuss our JCMT polarization observations and the previously observed \HII\ regions associated with NGC~6334 in Section~\ref{sec:obs}. In Section~\ref{sec:radial}, we discuss the methodology that we use to characterize the magnetic field morphologies of the \HII\ regions and discuss their impact on the cloud's magnetic field. In Section~\ref{sec:discussion}, we compare the magnetic pressure and tension with the  gas, radiation, and dynamic pressures in these regions and further explore techniques for identifying these \HII\ regions based on their dust polarization properties.  Section~\ref{sec:summary} summarizes our approach and findings.

\section{Observational data} 
\label{sec:obs}

In this section, we first discuss the JCMT SCUBA-2/POL-2  \citep{Bastien2011,Holland2013,Friberg2016} polarimetric observations of NGC~6334 at 850\,$\mu$m, taken as part of the B-fields In STar-forming Region Observations (BISTRO) survey \citep{Ward-Thompson2017}.  We then present the \HII\ regions of NGC~6334 as obtained from a variety of previously published catalogs. 

\subsection{JCMT BISTRO observations}

The 850\,$\mu$m dust polarized emission  observations were carried out under dry weather conditions with the atmospheric opacity ranging between $0.03$ and $0.07$ at 225\,GHz.
The data were reduced using the {\it pol2map}\footnote{\url{http://starlink.eao.hawaii.edu/docs/sc22.htx/sc22.html}} reduction code.  Observational and data reduction procedures are discussed in detail in Paper~\RN{1}, which also includes these data. 

\begin{figure}[hbpt]
\centering
\includegraphics[scale=0.4, trim={1.5cm 0.8cm 3cm 2.25cm}, clip]{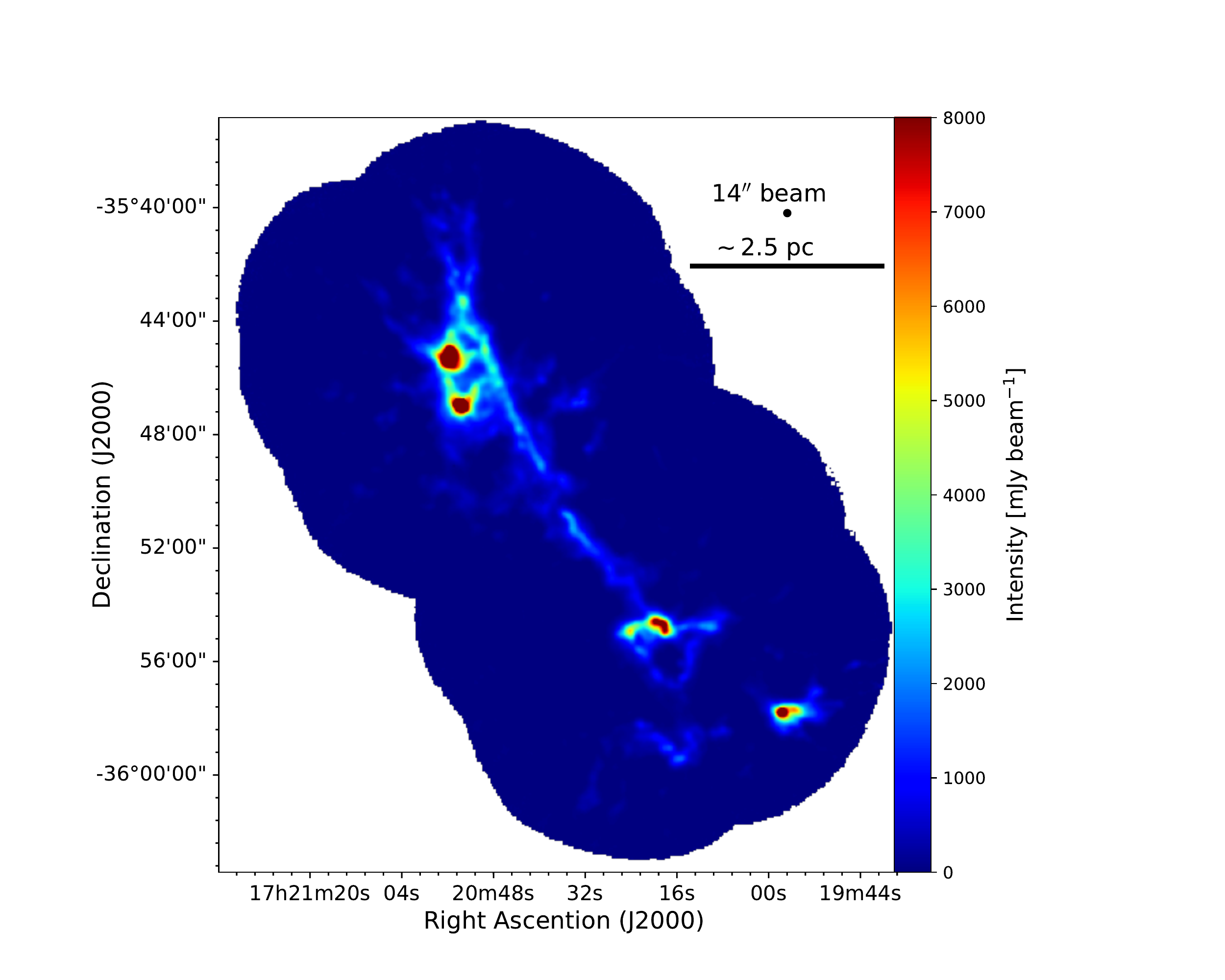}
\caption{NGC~6334 at 850\,$\mu$m with a pixel size of 4\arcsec.  The color image shows Stokes $I$ without any selection criteria applied. The cloud is composed of a main ridge ($\sim 10$\,pc long filament) along with dense clumps and cores. Evidence of \HII\ shells is also visible in the Stokes $I$ map.   }
\label{fig:NGC6334I}
\end{figure}

The observations have a spatial resolution of  $14\arcsec$ (Half Power Beam Width of the JCMT at 850\,$\mu$m). The maps of Stokes $I$ (the total dust thermal continuum emission), Stokes $Q$, and Stokes $U$  are re-projected onto Cartesian grids with pixel sizes of $4\arcsec$. 
The NGC~6334 molecular cloud is depicted in Figure~\ref{fig:NGC6334I} with pixel size of $4\arcsec$. The Stokes $I$, $Q$, and $U$ maps and their associated uncertainties are discussed in detail in Paper~\RN{1}.

We obtain the  polarization angle ($\psi$) and the initial polarized intensity ($PI_{\textsc{init}}$), using the Stokes parameters in the following equations:
\begin{equation}
\begin{aligned}
& \psi=0.5\,\arctan(\frac{U}{Q}),\\
& PI_{\textsc{init}} = \sqrt{Q^2+U^2}, 
\end{aligned}
\end{equation}
where the polarization angle is determined according to the IAU convention (from North to East in the equatorial coordinate system). The \bperp\ lines are perpendicular to the polarization lines, and their orientation ($\chi_{B_{\textsc{POS}}}$) is obtained using 
\begin{equation}
\chi_{B_{\textsc{POS}}}= \psi+90^\circ.
\label{psi}
\end{equation}

\begin{figure*}[hbpt]
\centering
\includegraphics[scale=0.9, trim={1.5cm 1cm 6.5cm 2cm} ,clip]{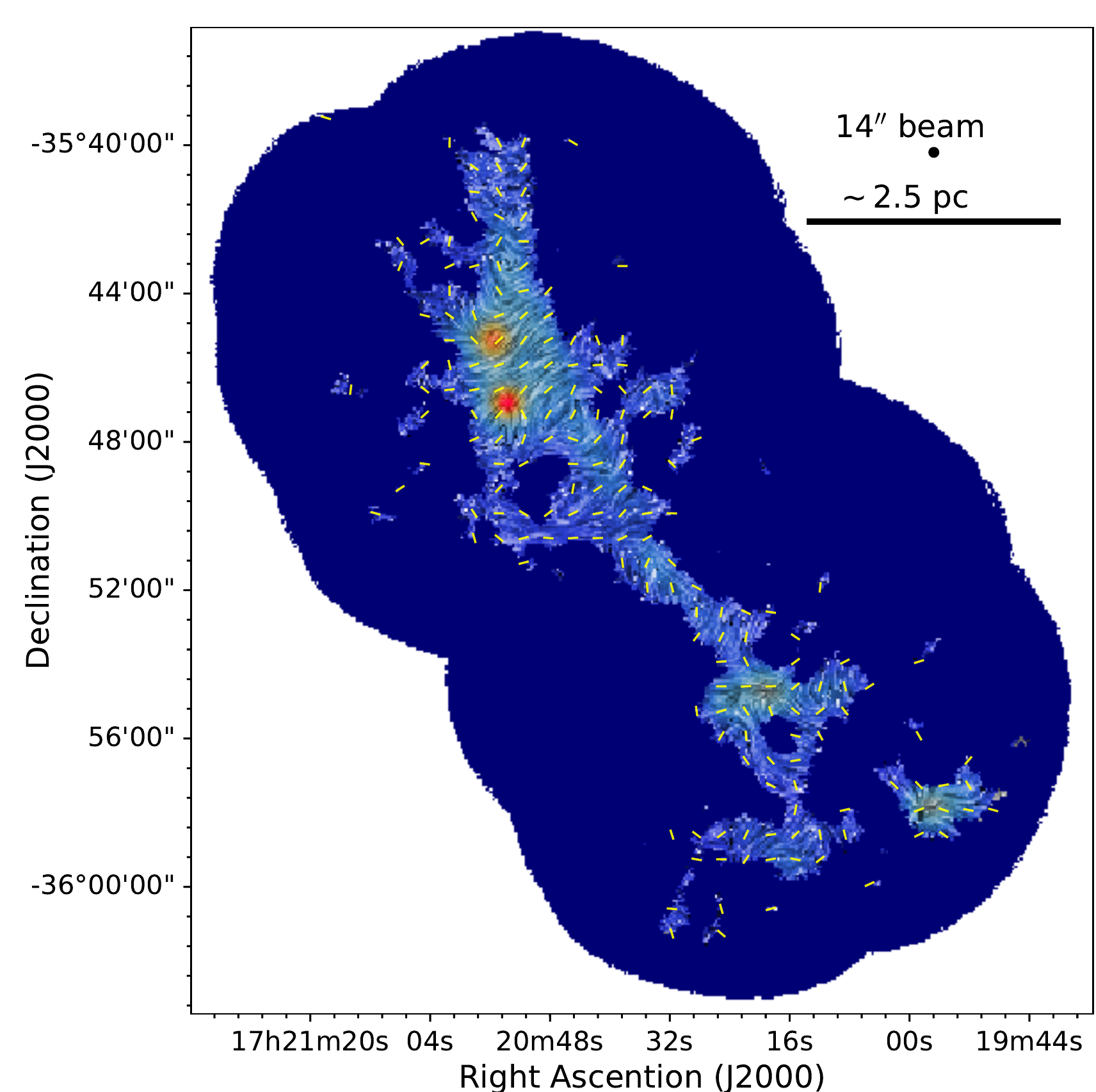}
\caption{Magnetic field lines associated with NGC~6334 at 850\,$\mu$m. The background color image illustrates the Stokes $I$ map and the drapery lines depict the plane-of-sky magnetic field lines in this region as obtained using the line integration convolution technique. The short yellow lines represent  the \bperp\ orientations (all of the same length), every ten pixels.  }
\label{fig:NGC6334LIC}
\end{figure*}

The parameter $PI_{\textsc{init}}$ is positively biased due to the squaring of the uncertainties in $Q$ (denoted by $\delta Q$) and $U$ (denoted by $\delta U$).  We debias  $PI_{\textsc{init}}$ and obtain the debiased polarization fraction ($PF$) as follows \citep[e.g.,][]{Serkowski1962, WardleKronberg1974}: 
\begin{equation}
\begin{aligned}
PI &= \sqrt{Q^2+U^2-0.5(\delta Q^2+\delta U^2)}, \\
PF &=\frac{PI}{I}.
\end{aligned}
\label{PIPF}
\end{equation}
This debiasing approach has been successfully used in previous studies for signal to noise ratios (SNR) higher than three  \citep[e.g.,][]{Vaillancourt2006,Plaszczynski2014,Montier2015,Hull2015,Pattleetal2019, Doietal2020, Pattleetal2021}. Subsequently, we find the uncertainties of the debiased polarization parameters as follows:
\begin{equation}
\begin{aligned}
&\delta PI =\frac{\sqrt{(Q\delta Q)^2+(U\delta U)^2}}{PI}, \\ 
&\delta PF = PF\, \sqrt{(\frac{ \delta PI}{PI})^2+(\frac{ \delta I}{I})^2}, \\
&\delta \psi  = 0.5 \frac{ \sqrt{(U\delta Q)^2+(Q\delta U)^2}}{PI^2},
\end{aligned}
\end{equation}
where the $\delta$ shows the uncertainty or the $\sqrt{variance}$ in each observed parameter. In this study, we use SNR($I$)$=\frac{{I}}{{\delta {I}}} >10$ and SNR($PI$)$=\frac{{PI}}{{\delta {PI}}} >3$ as data selection criteria, following Paper~\RN{1}. Figure~\ref{fig:NGC6334LIC} illustrates the overall magnetic field morphology of NGC~6334 using the line integration convolution technique~\citep[][]{LIC1993}. The background color image illustrates the Stokes $I$ map and the drapery lines depict the magnetic field lines. The short yellow lines represent the plane-of-sky magnetic field orientations,  every 10 pixels.

\subsection{Observed \HII\ regions}
\HII\ regions are best identified and confirmed using Radio Recombination Lines (RRL) or \Halpha\ observations. However, surveys by \citet{Baniaetal2010} and \citet{Andersonetal2011} demonstrated that MIR observations can be  used to identify these regions, which appear as a bubble-like structure in $\sim 22\,\mu$m  observations surrounded by a shell-like structure at $\sim 12\,\mu$m wavelength. Traditionally the \HII\ regions in NGC~6334 were named A to E when observed in radio \citep{Rodriguezetal1982}, and \RN{1} to \RN{5} when observed in far infrared ~\citep[FIR;][]{McBreenetal1979}. For a complete list of all observed \HII\ regions in NGC~6334, we use the Wide-field Infrared Survey Explorer~\citep[WISE\footnote{http://astro.phys.wvu.edu/wise/};][]{Andersonetal2014} and \citet{Simpsonetal2012} catalogs (observed in MIR).  We refer to the \HII\ regions identified in the WISE and \citet[][]{Simpsonetal2012} catalogs as WISE and Simpson bubbles, respectively. 

The WISE catalog compiles more than 8000 confirmed and candidate \HII\ regions in the Galactic disk that were observed by the WISE satellite in the 12$\,\mu$m and 22$\,\mu$m bands~\citep[][]{Andersonetal2014, Andersonetal2015}. The \HII\ bubbles in the WISE catalog are categorized as Known (K), Groups (G), Candidates (C), and radio Quiet candidates (Q). Known regions are those that have been observed in either RRL or \Halpha . When Candidate \HII\ regions are spatially associated with known \HII\ regions they are referred to as Groups. 
A radio Quiet candidate is a region that is normally undetectable by radio continuum surveys due to their low sensitivity, but which is detected by WISE (with the 6\,mJy sensitivity of WISE at 22$\,\mu$m).
Candidates (C and Q) are typically found to be \HII\ regions 95\% of the time~\citep{Andersonetal2014, Andersonetal2015, Andersonetal2018}.

Additionally, we use the \HII\ catalog from \citet{Simpsonetal2012} to further locate regions in NGC~6334. 
The bubbles identified by \citet{Simpsonetal2012} are part of the citizen science Milky Way Project\footnote{http://www.milkywayproject.org} (MWP), in which online volunteers determined the location of the bubbles using Spitzer observations in the 8$\,\mu$m and 24$\,\mu$m bands. The WISE and Simpson catalogs identify \HII\ regions with slightly different centers and radii. This is likely because these bubbles are not perfectly spherical, and the two studies use different methods to determine the radii and shell thickness of \HII\ regions (based on their MIR characteristic morphology): \citet[][WISE survey]{Andersonetal2014} use circles to identify the regions, whereas \citet[][Simpson catalog]{Simpsonetal2012} rely on volunteer assistance to associate elliptical objects with these regions (after which the regions are chosen based on their selection number\footnote{hit rate} and an effective radius is determined to describe the radius of each one).

\begin{figure*}[thbp]
\centering
\includegraphics[scale=0.83,clip,trim={1cm 1cm 4cm 2cm}]{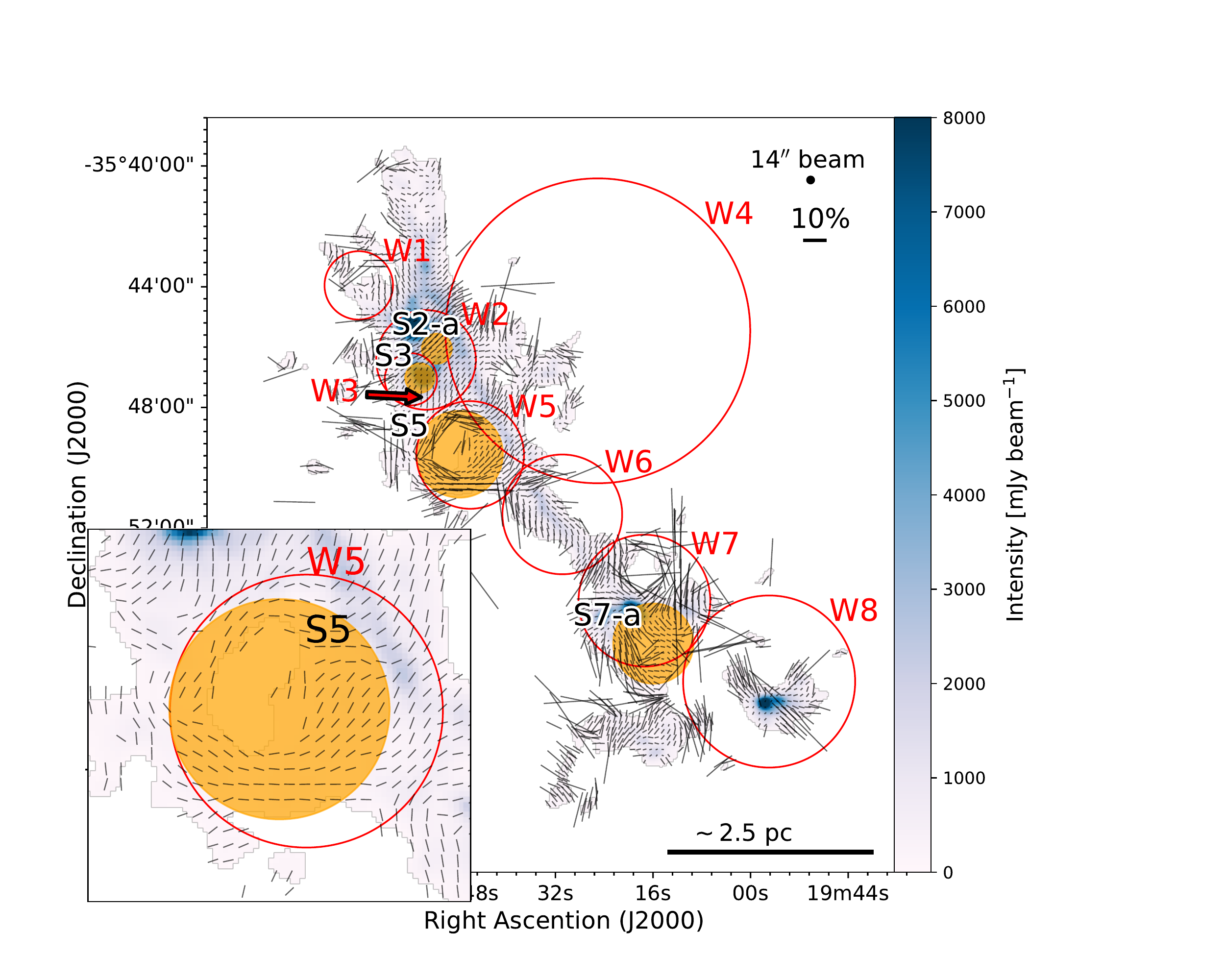}
\caption{\HII\ regions associated with NGC~6334.  The background color image shows the Stokes $I$ map. The black lines represent the plane-of-sky magnetic field orientation for a $12\arcsec$ pixel-sized map (in order to make it appear less crowded visually). The length of each field line is proportional to the polarization fraction. The red circles depict the WISE bubbles, which are numbered with a red font on their top-right side.  The orange disks represent the Simpson bubbles, which are labeled in black font on their upper-left corner. The inset shows a zoomed-in view of W5 (and S5), which clearly exhibits tangential field (radial polarization) lines as equal-length segments relative to the bubble.}
\label{fig:NGC6334Bubbles}
\end{figure*}

The WISE and Simpson bubbles associated with our observations are illustrated in  Figure~\ref{fig:NGC6334Bubbles} and listed in Tables~\ref{table:Wise} and \ref{table:Simpson}. The WISE bubbles are numbered W1 to W8, depicted with red circles in Figure~\ref{fig:NGC6334Bubbles}. The Simpson bubbles are denoted by the letter S and are numbered similarly to the corresponding WISE bubbles; if a WISE and a Simpson bubble depict a  similar bubble in the sky, they are numbered identically (for example, W5 and S5); if a Simpson bubble is  nested within a WISE bubble (at a similar location but with a significantly smaller radius), it is numbered identically as the WISE bubble but with the suffix ``a'' (for example W7 and S7-a).

Table~\ref{table:Wise} lists the WISE bubbles along with their equatorial coordinates, radius, distance \citep[parallax measurements by][]{Wuetal2014}, category, and the local standard of rest velocity (km\,s$^{-1}$). We note that the mean molecular and ionized gas velocity of NGC~6334 are $-4$\,km\,s$^{-1}$\citep{Russeiletal2016} and $-3$\,km\,s$^{-1}$\citep{CaswelHaynes1987}, respectively. Bubbles W4 to W8 are ``Known'' bubbles, identified by \citet{Quirezaetal2006} using Carbon RRL (C{\sc ii}) at 3.5\,cm. The physical scale of Bubble W4 is comparable to that of the NGC 6334 cloud, and it likely contributed to the formation of NGC 6334
~\citep[][]{Fukuietal2018PASJ}; consequently, it is not extensively discussed in our analysis of \HII\ regions within the cloud.
The Simpson bubbles are listed in Table~\ref{table:Simpson} along with their equatorial, effective radius (R$_{\text{eff}}$), effective thickness (Th$_{\text{eff}}$), inner and outer diameters (along the $\delta$ and $\alpha$ axes), eccentricity, and position angle of the ellipsoid bubble. 
Each of the Flags 1, 2, and 0 denotes a distinct bubble type. Flag 1 denotes bubbles with smaller bubbles on their edges, Flag 2 represents those found within larger bubbles, and Flag 0 indicates neither of these two.

\begin{table*}[thbp]
\centering
 \begin{tabular}{| c c c c c c c c|} 
 \hline
Number	&   Name	&   $\alpha$ ($^{\circ}$ J2000)	&   $\delta$ ($^{\circ}$ J2000)	&   Radius ($\arcsec$) & Dist (kpc) 	& Category & V$_{LSR}$ (km\,s$^{-1}$) \\ \hline \hline
W1	    &	G351.479+0.643	&	260.268	&	$-35.734$	&	68.0    &	--	&	Q   &   --      \\
W2	    &	G351.424+0.65	&	260.223	&	$-35.775$	&	99.0    &	--	&	G   &   --      \\
W3	    &	G351.42+0.637	&	260.233	&	$-35.786$	&	52.0    &	--	&	Q   &   --      \\
W4	    &	G351.383+0.737	&	260.106	&	$-35.759$	&	303.0   &	1.3	&	K   &   $-3.4$  \\
W5	    &	G351.367+0.64	&	260.193	&	$-35.828$	&	107.0   &	1.3	&	K   &   $-3.4$  \\
W6	    &	G351.311+0.663	&	260.13	&	$-35.86	$   &	119.0   &	1.3	&	K   &   $-3.4$  \\
W7	    &	G351.246+0.673	&	260.074	&	$-35.908$	&	131.0   &	1.3	&	K   &   0.6     \\
W8	    &	G351.17+0.704	&	259.989	&	$-35.953$	&	171.0   &	--	&	K   &   0.5     \\    
 \hline
 \end{tabular}
 \caption{Bubbles found in the WISE catalog associated with NGC~6334. The name, right ascension ($\alpha$), and declination ($\delta$) for the center of each bubble are indicated.  The velocities (V$_{LSR}$) are obtained from~\citet{Quirezaetal2006} and distances (dist) from \citet{Wuetal2014} using parallax measurements. The bubble categories are Known (K), radio Quiet (Q), and Group (G), as identified in the WISE catalog.}
 \label{table:Wise}
\end{table*}

\begin{table*}[thbp]
\centering
 \begin{tabular}{| c c c c c c c c c c c c  |} 
 \hline
Number  &  $\alpha$($^{\circ}$ J2000)	 & $\delta$($^{\circ}$ J2000)   & R$_{\text{eff}}$ ($'$)	& Th$_{\text{eff}}$ ($'$)	& InnXDia ($'$) & InnYDia ($'$)   & OutXDia ($'$)	& $\epsilon$     & PA($^{\circ}$) & Hit	    &    Flag        \\\hline \hline
S2-a      &  260.214	                     & $ -35.768$                   & 0.50	            & 0.61	            & 0.630   & 0.871	  & 1.240	& 0.691	         & 97	          & 0.16	&      0     \\
S3      &  260.225	                     & $ -35.784$                   & 0.49	            & 0.61	            & 0.728   & 0.730	  & 1.335	& 0.068	         & 10	          & 0.22	&      0     \\
S5     &  260.199	                     & $ -35.826$                   & 1.44	            & 1.46	            & 2.212   & 2.262	  & 3.676	& 0.208	         & 33	          & 0.16	&      1     \\
S7-a     &  260.067	                     & $ -35.931$                   & 1.33	            & 1.64	            & 1.936   & 1.986	  & 3.574	& 0.222	         & 37	          & 0.11	&      0     \\
 \hline
 \end{tabular}
 \caption{Bubbles identified by \citet{Simpsonetal2012} associated with NGC~6334. The center of each bubble is shown in equatorial coordinates with $\alpha$ and $\delta$. R$_{\text{eff}}$	and Th$_{\text{eff}}$ represent the effective radius and thickness associated with each bubble. The bubbles are first identified as elliptical shapes (before determination of R$_{\text{eff}}$	and Th$_{\text{eff}}$) with inner diameters of InnXDia and InnYDia (along the $\delta$ and $\alpha$ axes), outer diameter of OutXDia (along the $\delta$ axis), eccentricity of $\epsilon$, and ellipse position angle of PA. The ``hit rate'' of a bubble represents the level of agreement among the Milky Way project users regarding bubble detection and is  the ratio of the number of bubbles drawn that qualify as an \HII\ region to the number of times the bubble was detected by the users.
 The final catalog includes only bubbles with a hit rate of 0.1 or greater. Flags 1, 2, and 0 indicate bubbles: with smaller bubbles on their edge, positionally located within a larger bubble, and neither of the two, respectively.}
\label{table:Simpson}
\end{table*}

\section{\HII\ magnetic field morphologies}
\label{sec:radial}

\begin{figure}[thbp]
\centering
\includegraphics[scale=0.33,clip, trim={0cm 0cm 0cm 0.5cm}]{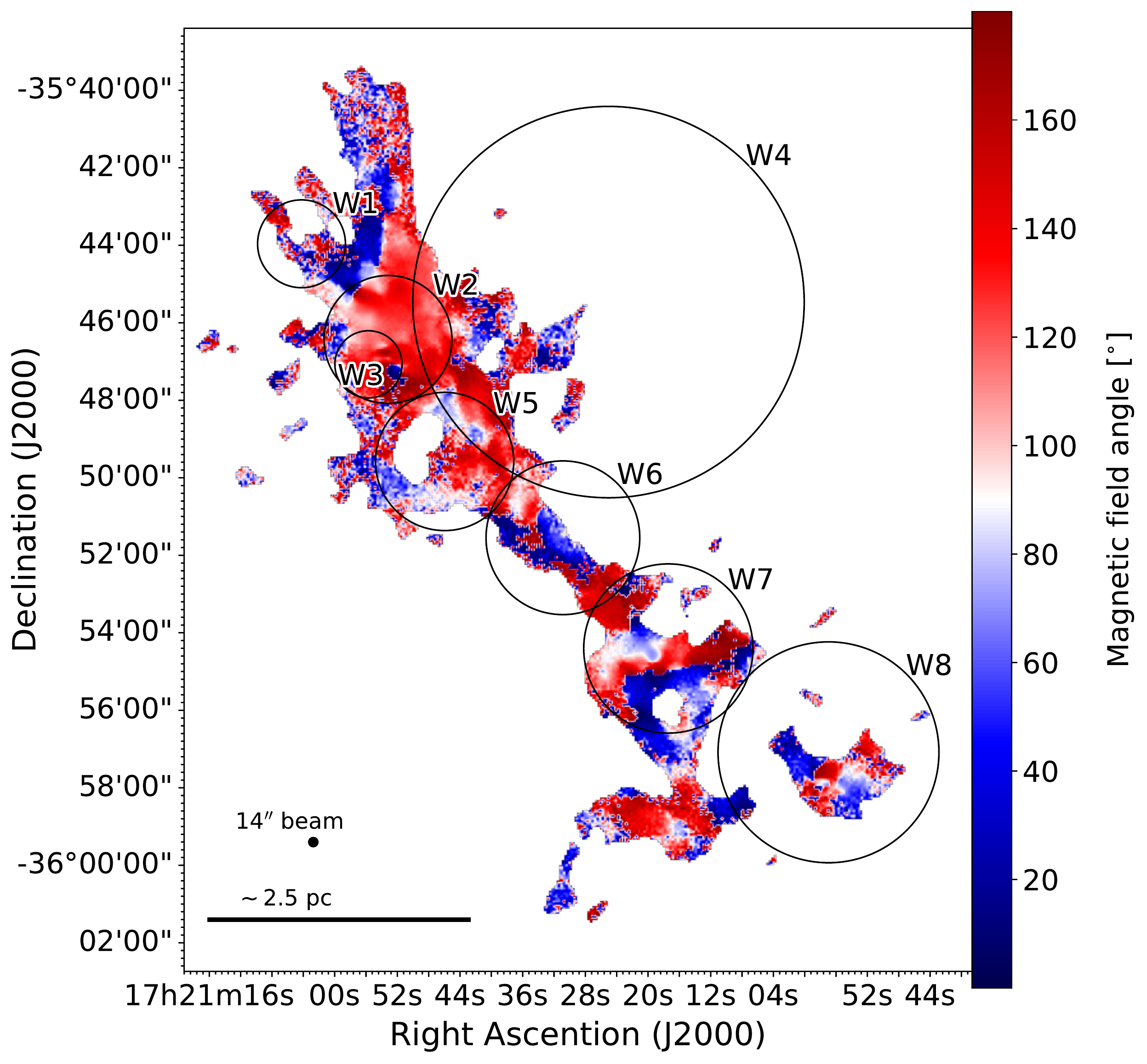}
\caption{Plane-of-sky magnetic field angles ($\chi_{B_{\textsc{POS}}}$) in NGC~6334. The background color image shows the magnetic field angles  in degree. The angles are in the IAU convention system. The magnetic field angles surrounding some of the bubbles (particularly W5) indicate wrapping of the field lines 
(changing from blue to red to blue).  WISE bubbles are shown with black circles.}
\label{fig:AngleMap}
\end{figure}

\HII\ regions can push away their surrounding interstellar medium (and the frozen-in field lines),  resulting in tangential magnetic field (radial polarization\footnote{We use the terms radial polarization and tangential magnetic field interchangeably in this study, as the field lines are perpendicular to the polarization lines.}) patterns. Moreover, the magnetic field lines surrounding these bubbles are likely to have been compressed \citep[e.g.,][]{Eswaraiahetal2020} and ordered, potentially resulting in increased polarization fractions. These effects can be seen in our observations. The plane-of-sky magnetic field lines depicted in Figure~\ref{fig:NGC6334Bubbles} indicate tangential fields (radial polarization) associated with some of these \HII\ regions. This field morphology is evident in W5 (S5), as illustrated in the inset of Figure~\ref{fig:NGC6334Bubbles}, representing equally sized magnetic field lines in a zoomed-in view of this bubble. This wrapping of the field lines appears as a periodic polarization angle pattern along the circumference of each bubble, resulting in contiguous areas of below-90$^{\circ}$ and above-90$^{\circ}$ magnetic field angles surrounding each bubble. Figure~\ref{fig:AngleMap} illustrates signatures of this behavior in $\chi_{B_{\textsc{POS}}}$ (cyclic orientations of \bperp\ lines changing from blue to red to blue to red in the figure), which is characteristic of tangential field morphologies (wrapping of field lines around a bubble).  In this figure, blue denotes a magnetic field angle less than 90$^{\circ}$ and the red color shows a magnetic field angle greater than 90$^{\circ}$ (in the IAU convention). 

\begin{figure}[thbp]
\centering
\includegraphics[scale=0.6,clip, trim={0cm 1cm 0cm 1cm}]{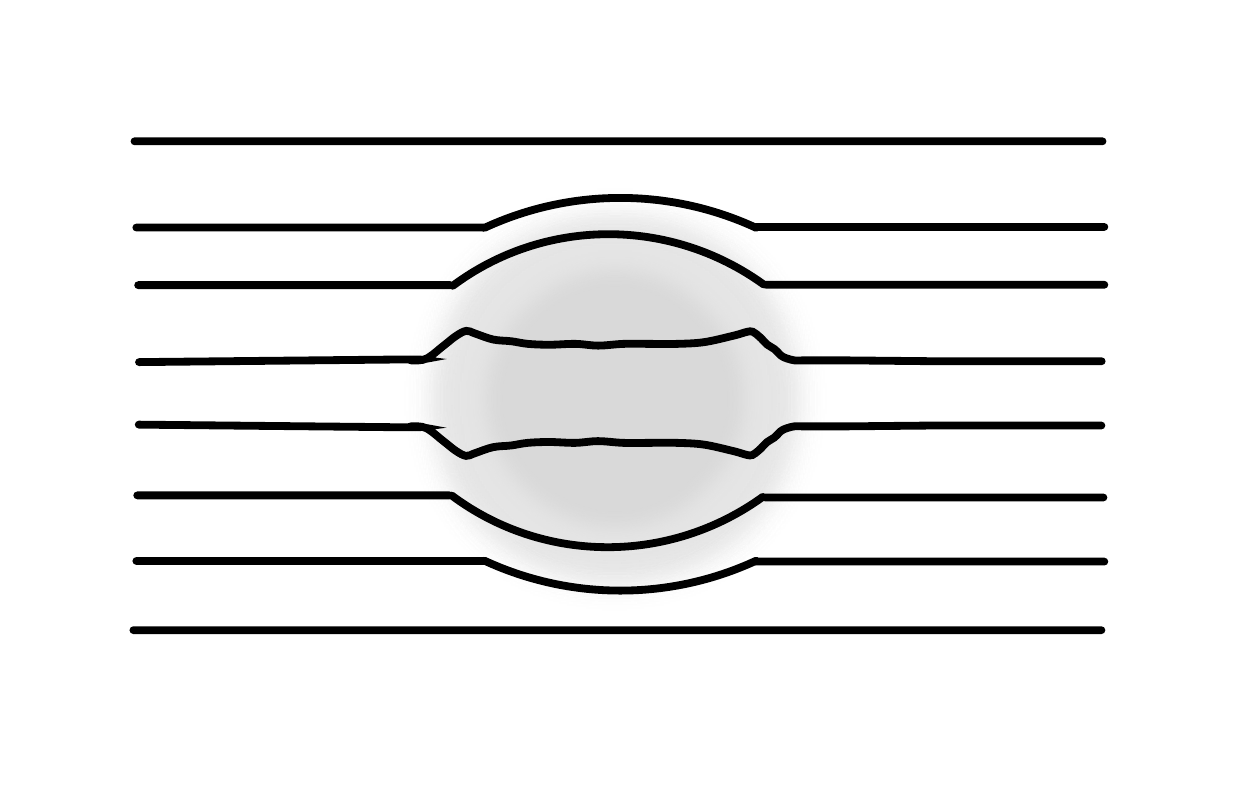}
\caption{Co-evolution of magnetic field lines and \HII\ regions as studied by \citet[][]{Krumholzetal2007}. The black lines and the gray circle show the magnetic field lines and the \HII\ region, respectively. The field lines are pushed by the \HII\ region, resulting in tangential field patterns.}
\label{fig:HIICartoon}
\end{figure}

Furthermore, theoretical studies~\citep[e.g.,][]{Krumholzetal2007} demonstrate that \HII\ regions can influence and alter the magnetic field morphology of their parental molecular cloud. \citet[][]{Krumholzetal2007} simulate the evolution of magnetized \HII\ regions, resulting in tangential magnetic field lines. Figure~\ref{fig:HIICartoon} depicts their resulting field morphology after approximately 0.53\,Myr with an initial magnetic field strength of $14.2\,\mu$G.  We examine the \HII\ magnetic field morphologies and their polarization fractions in this section.

\subsection{Identifying field morphologies: Methodology}
To investigate the polarization patterns associated with the \HII\ bubbles and determine whether they represent a tangential magnetic field pattern, we transform the polarization frame of reference for each bubble individually and define radial Stokes parameters $Q_r$ and $U_r$. In this reference frame, a polarization line along the radius of the bubble has a zero polarization angle. 
At each position $(\alpha, \delta)$ about the bubble center $(\alpha_0, \delta_0)$, we obtain $Q_r$ and $U_r$ using:  
\begin{equation}
\begin{aligned}
Q_r &= +Q\cos{2\phi} +U\sin{2\phi}, \\
U_r &= -Q\sin2\phi + U\cos 2\phi , 
\label{eq:qrur}
\end{aligned}
\end{equation}
where $\phi = \arctan{\frac{\alpha-\alpha_0}{\delta-\delta_0}}$ is the polar angle of a given position  with respect to  the center of the bubble in equatorial coordinates.  The angle $\phi$ is zero when pointing toward North and increases toward East. We then determine the average $Q_r$ and $U_r$ at fixed distances from the bubble center (i.e., radial profile of $<Q_r>$ and $<U_r>$, averaged within thin, two-pixel-wide annuli around the bubble). A positive $<Q_r>$ value indicates that a shell demonstrates radial polarization pattern on average, whereas a negative value hints to tangential polarization. $U_r$ represents polarization at a $\pm 45^{\circ}$ angle to the radial direction. Therefore, $<U_r> = 0$  indicates radial or tangential polarization at that radius. 

This formalism  enables us to more easily identify radial polarization (tangential  magnetic field patterns) associated with individual bubbles.  \citet{Schmidetal2006} previously employed this technique to study the radial polarization associated with Uranus and Neptune. They found that the technique was particularly effective in reducing uncertainties caused by systematic errors in the data reduction process.  Additionally, \citet{Canovasetal2015} used the technique to investigate the polarization of protoplanetary disks. 

To study $<Q_r>$ and $<U_r>$ more efficiently and to determine the radial polarization patterns, we define $\theta_r$ as follows:
\begin{equation}
   \theta_r = 0.5 \times \arctan\big(\frac{<U_r>}{ <Q_r>}\big). 
   \label{eq:theta}
\end{equation}
If a shell exhibits radial polarization on average (with positive $<Q_r>$ and near zero $<U_r>$ values), then $\cos(\theta_r)$ should be close to 1.  We look for radii with $\cos(\theta_r) > 0.95$ (allowing for 5\% total variation or uncertainty) including the error bars, to identify shells with radial polarization patterns. Appendix~\ref{apndx:radialPolErrorBar} contains a discussion of the method used to calculate the error bars.

\subsection{Polarization pattern of \HII\ regions in NGC~6334: Results}

\begin{figure*}[thbp]
\centering
  \vspace{0.2cm}
  \centering\small (a) Bubble W1, WISE category Q. 
  \begin{tabular}{c c}
    \includegraphics[scale=0.4, trim={0.cm 0cm 0cm 0cm} ,clip]{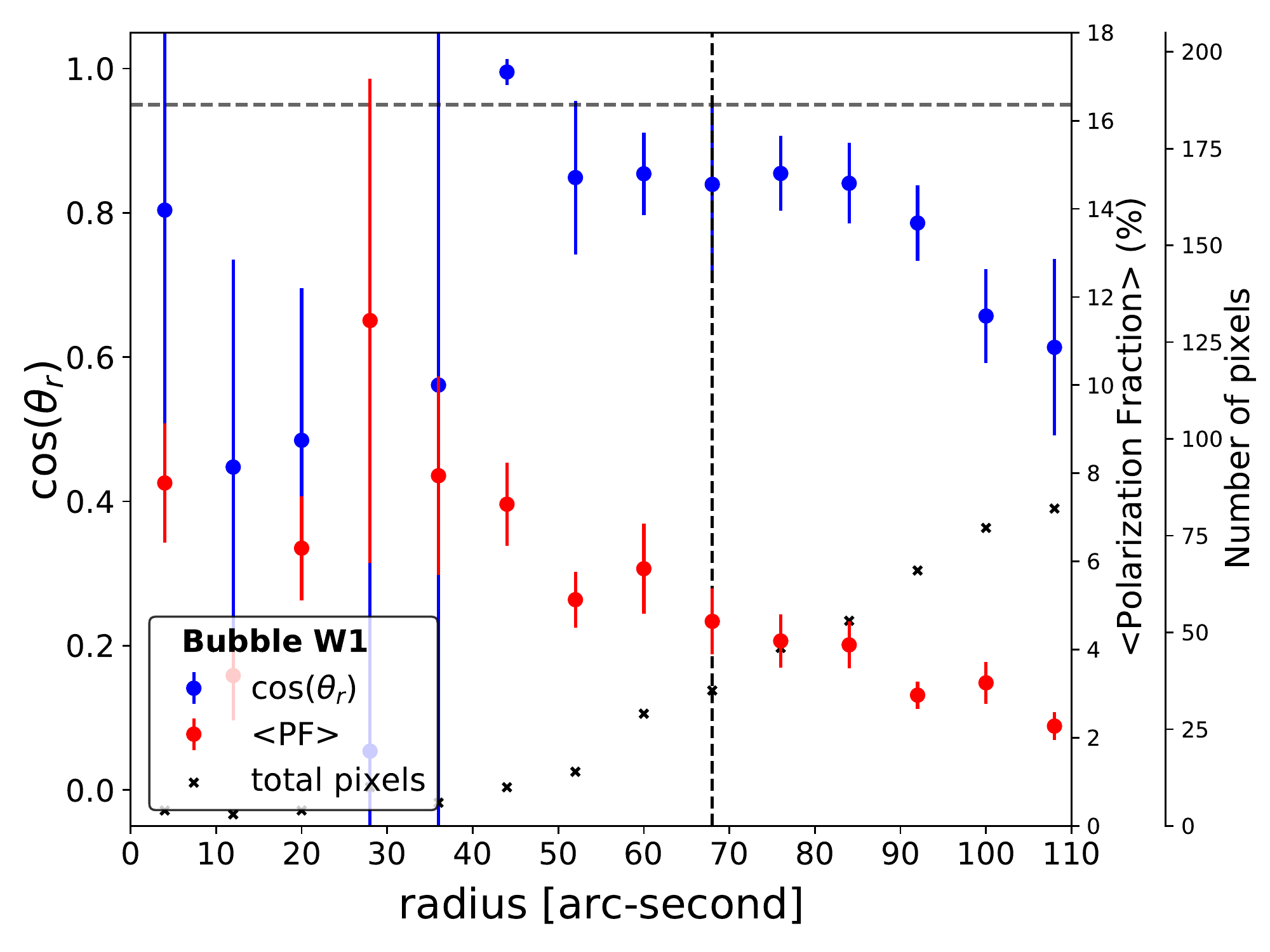}
    \includegraphics[scale=0.4, trim={0cm 0cm 1cm 0cm} ,clip]{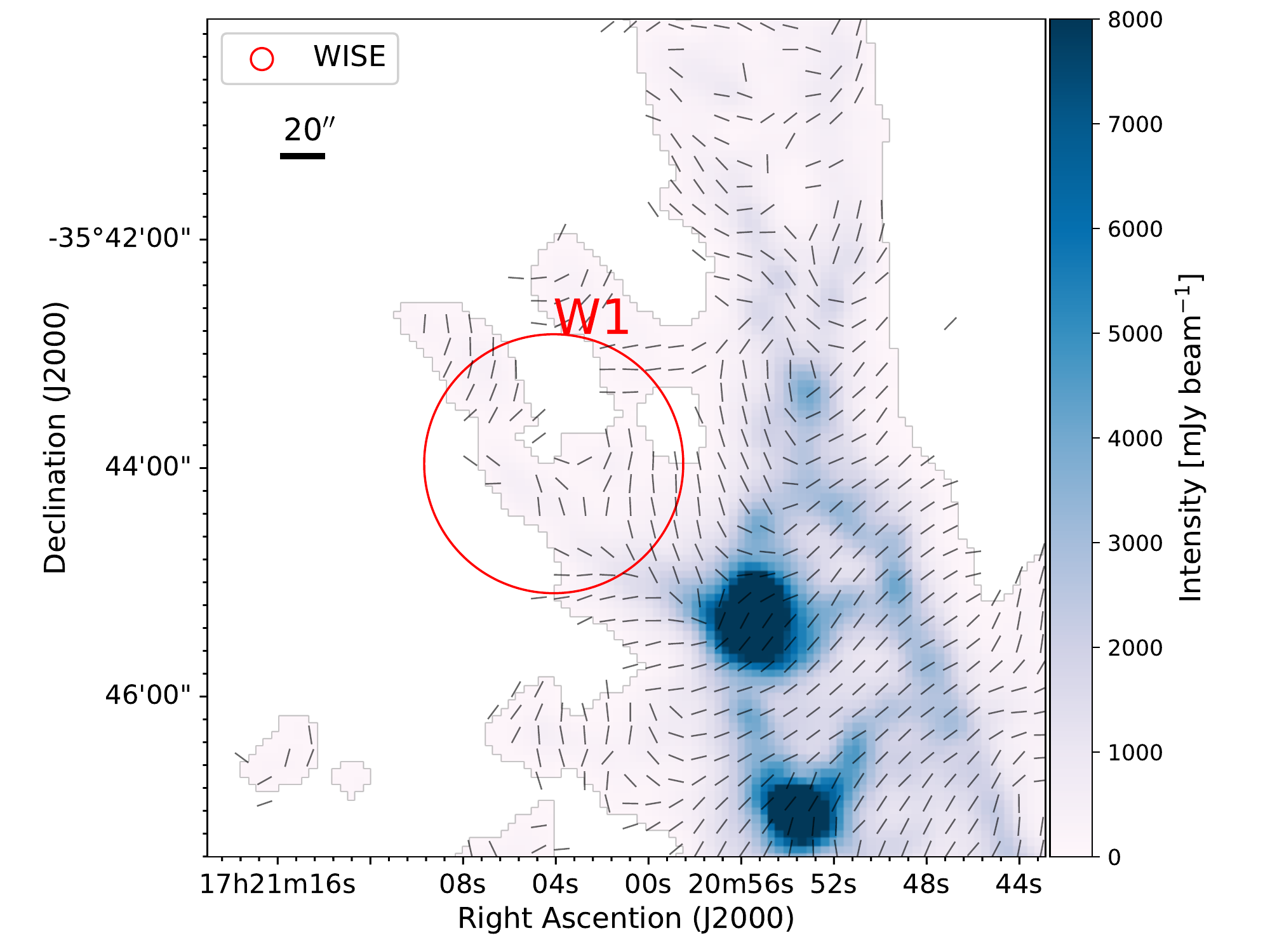}\\
    \vspace{0.05cm}
  \end{tabular}%
   \\ 
   \centering\small (b) Bubble W6, WISE category K.
  \begin{tabular}{c c}
    \includegraphics[scale=0.4, trim={0.cm 0cm 0cm 0cm},clip]{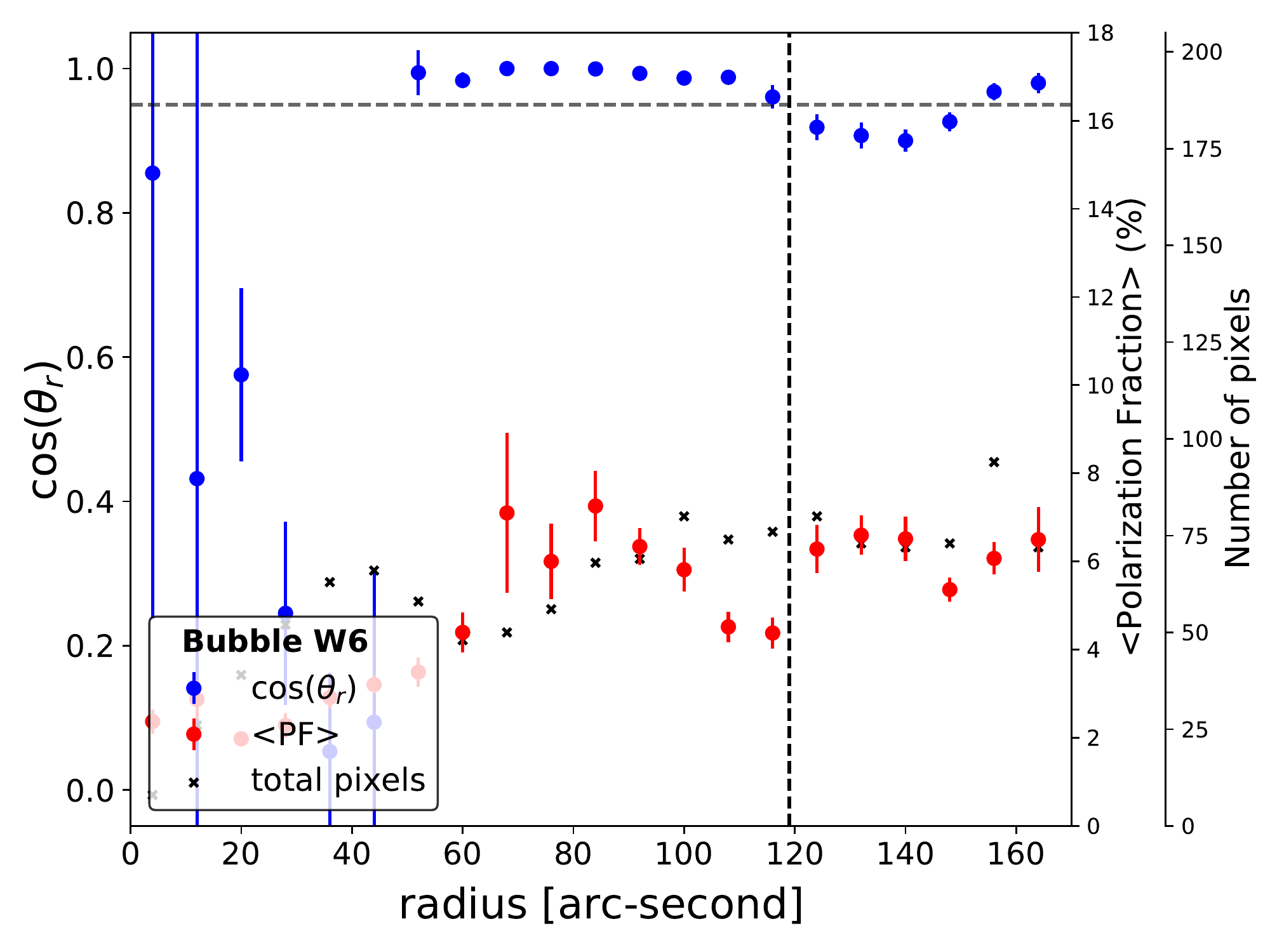}
    \includegraphics[scale=0.4, trim={0cm 0cm 1cm 0cm},clip]{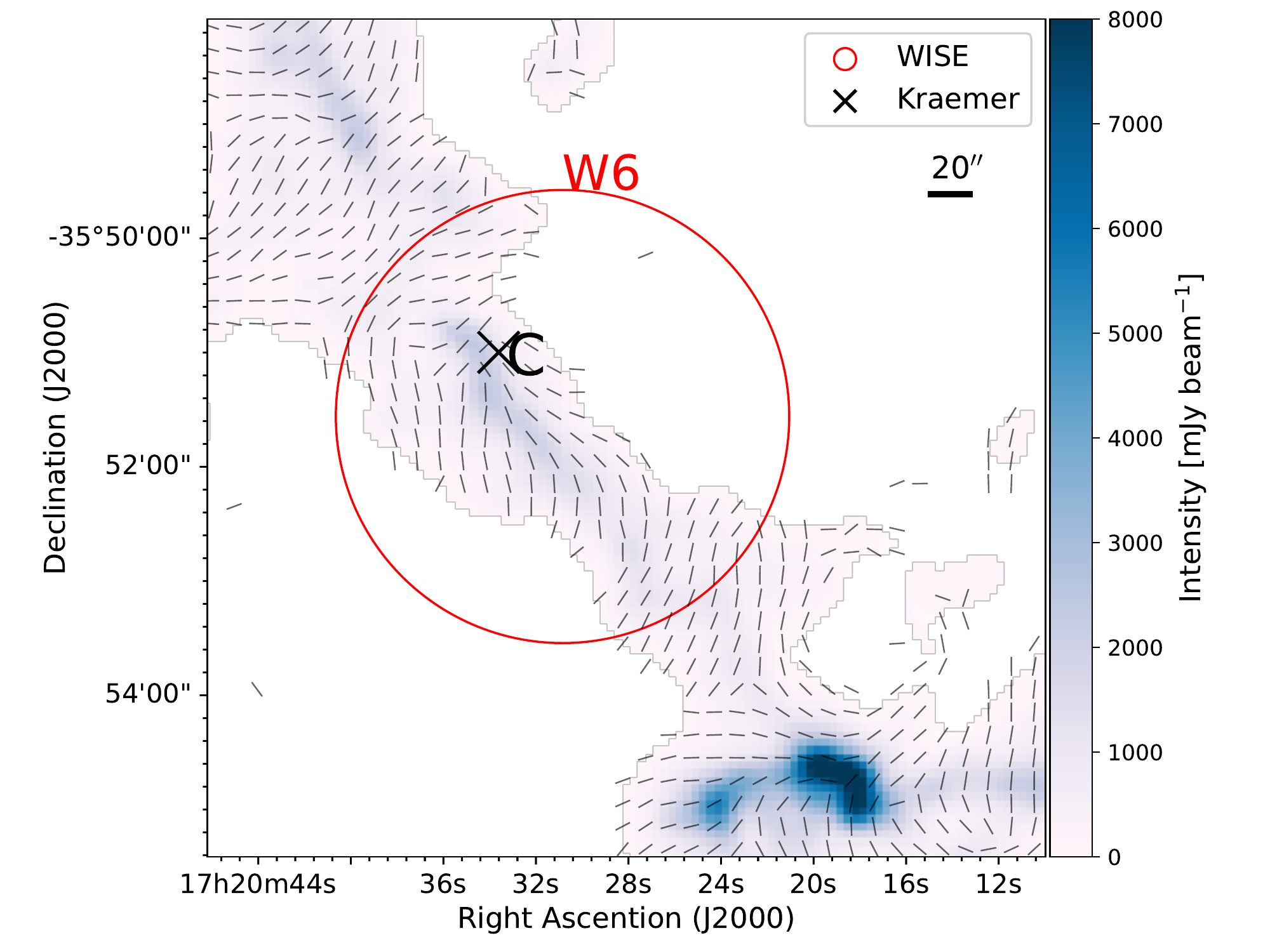}\\
    \vspace{0.05cm}
  \end{tabular}%
  \\ 
  \centering\small (c) Bubble W8, WISE category K.
  \begin{tabular}{c c}
    \includegraphics[scale=0.4, trim={0.cm 0cm 0cm 0cm},clip]{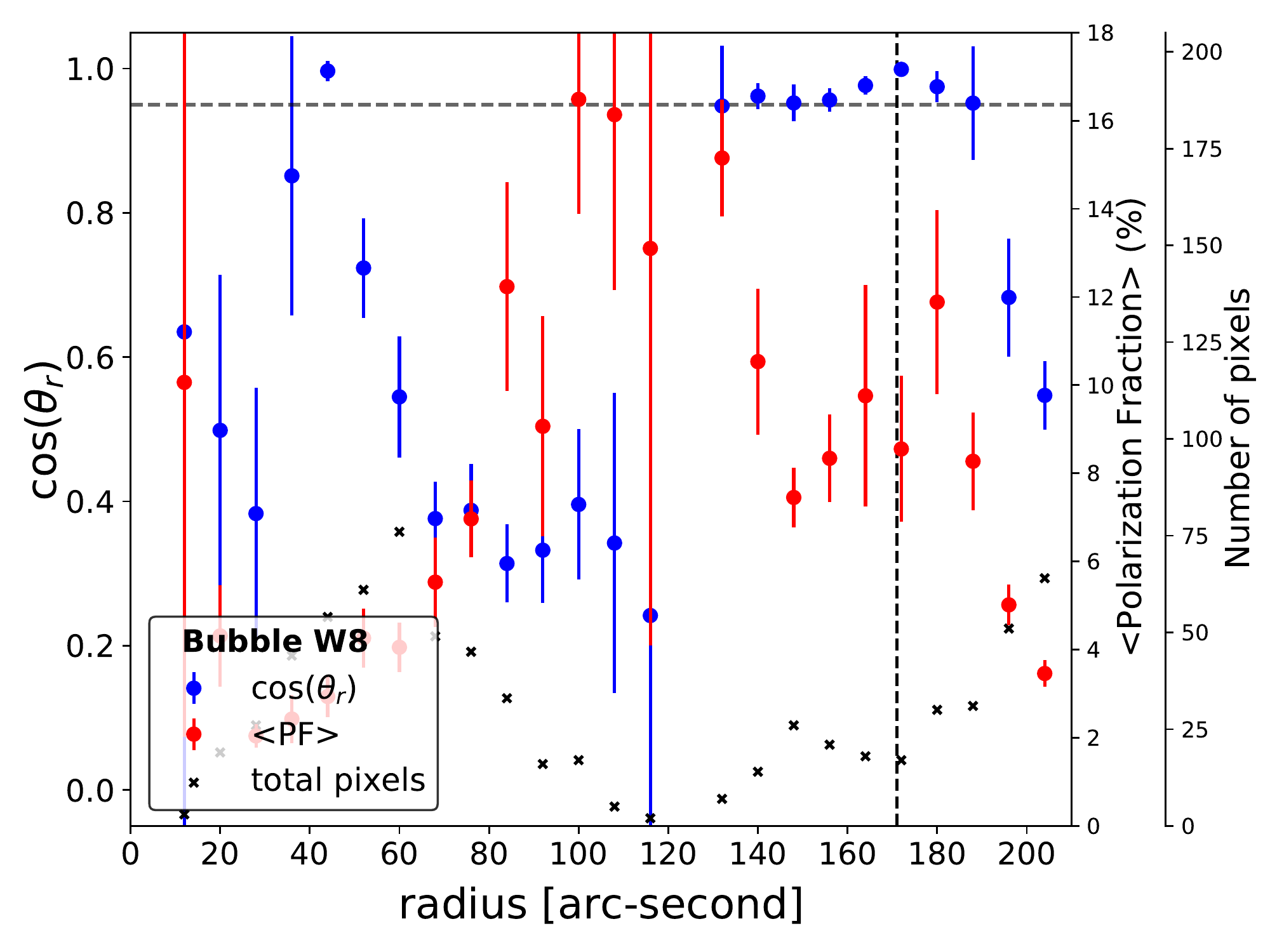}
    \includegraphics[scale=0.4, trim={0cm 0cm 1cm 0cm},clip]{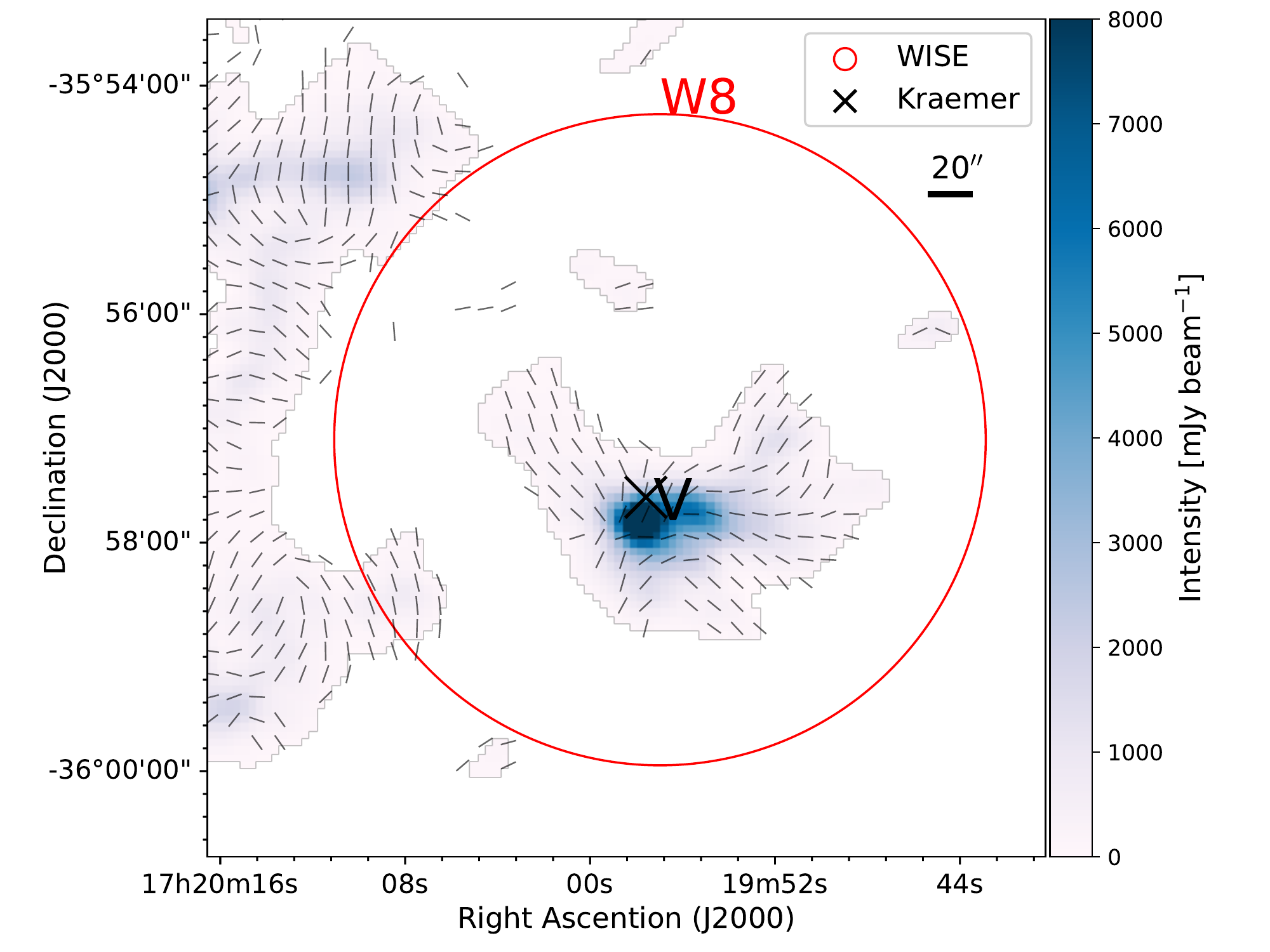}\\
  \end{tabular}%
\caption{Radial polarization in W1, W6, and W8. \textbf{Left column}: Radial profile of $\cos(\theta_r)$ and the mean polarization fraction in each bubble.  The dashed vertical and horizontal lines indicate the radius of each bubble and  $\cos(\theta_r) = 0.95$, respectively. The blue, red, and black markers indicate  $\cos(\theta_r)$, the mean polarization fraction, and  the total number of pixels (which satisfy the selection criteria of SNR($I$)$ >10$ and SNR($PI$)$> 3$) in each shell, respectively. The error bars in polarization fraction represent the standard deviation of the mean. The error bars for the blue markers are discussed in Appendix~\ref{apndx:radialPolErrorBar}.  \textbf{Right column:} Zoomed-in view of each bubble. The black lines and the background color image represent equal-sized magnetic field lines and Stokes $I$, respectively. }
\label{fig:RadialPol2column}
\end{figure*}

\begin{figure*}[thbp]
\centering
\vspace{0.2cm}
  \centering\small (a) Bubble W5, WISE category K.
  \begin{tabular}{c c c}
    \includegraphics[scale=0.3, trim={0cm 0cm 0cm 0cm},clip]{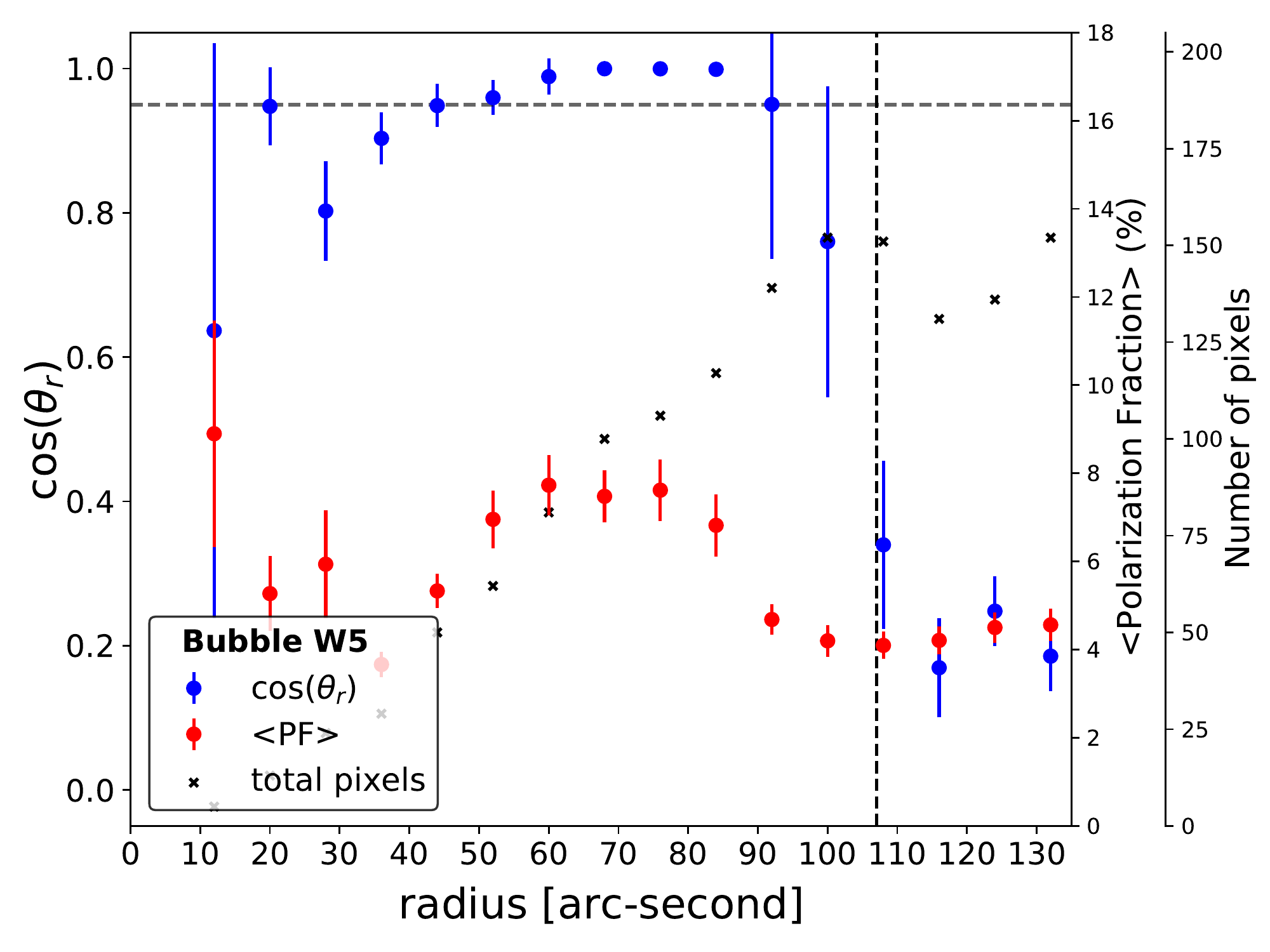}
    \includegraphics[scale=0.3, trim={0cm 0cm 0cm 0cm},clip]{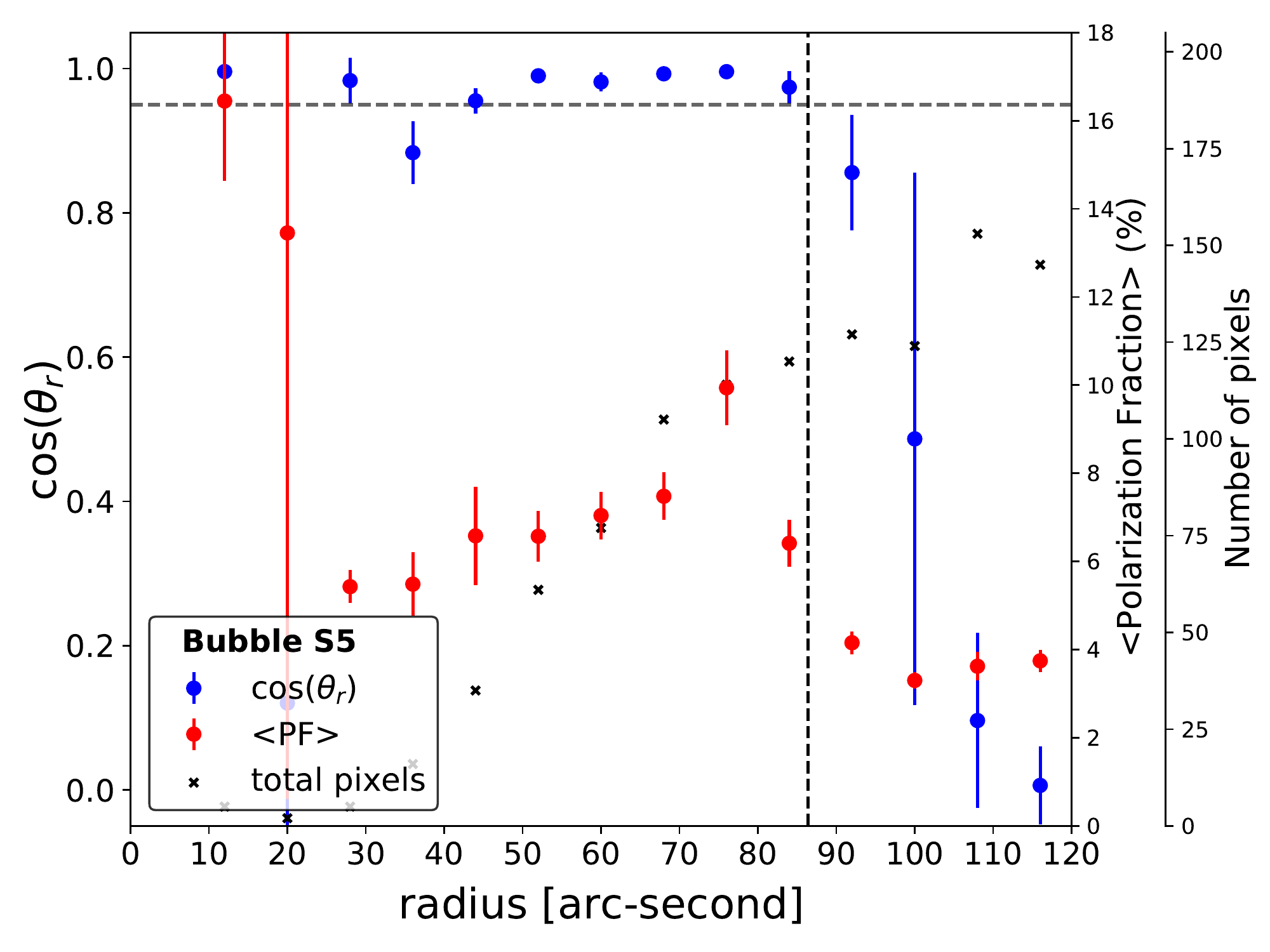}
    \includegraphics[scale=0.3, trim={0cm 0cm 1cm 0cm},clip]{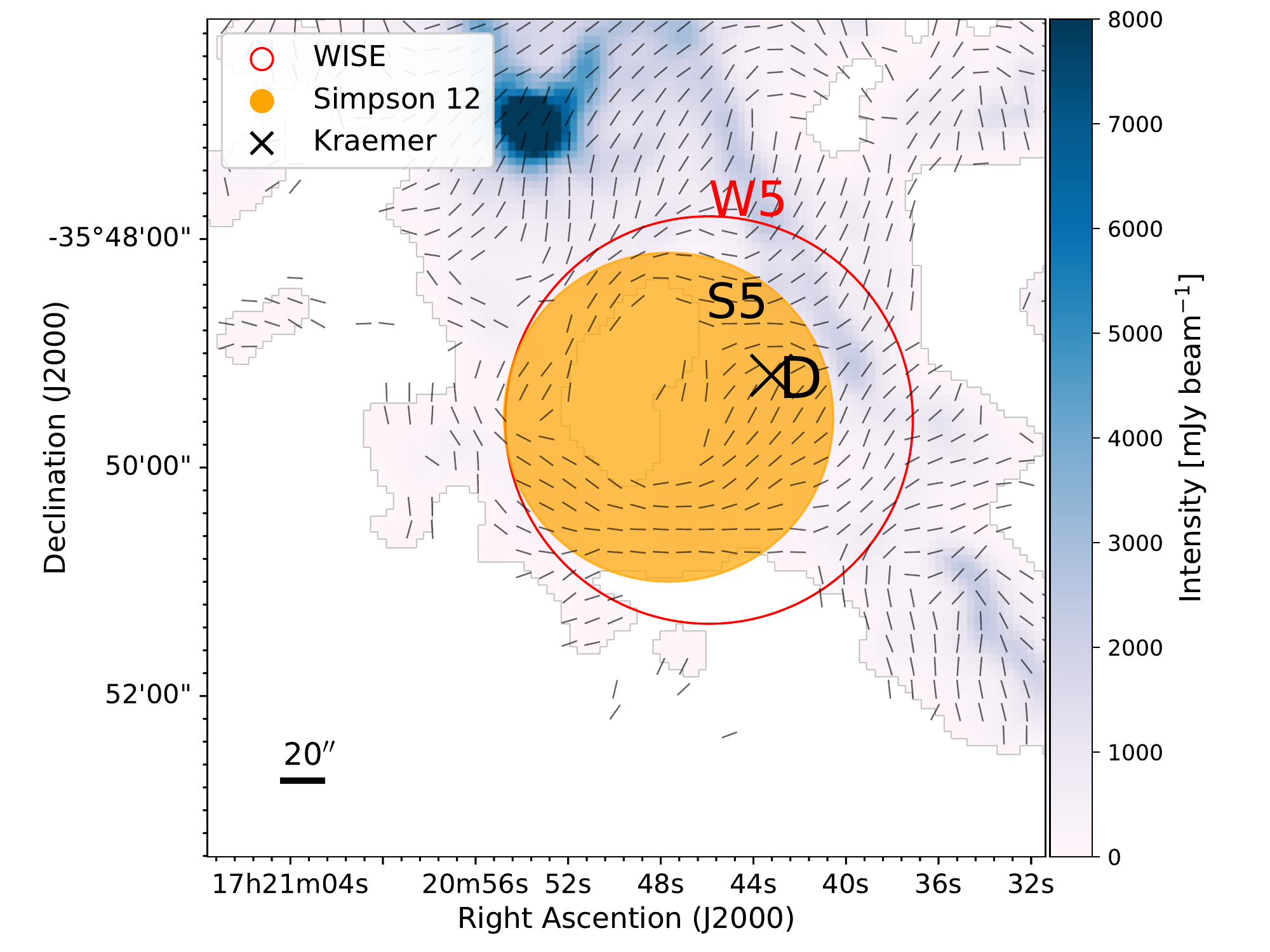}\\
    \vspace{0.05cm}
  \end{tabular}%
 \\ 
 \centering\small (b) Bubble W7, WISE category K.
  \begin{tabular}{c c c}
    \includegraphics[scale=0.3, trim={0cm 0cm 0cm 0cm},clip]{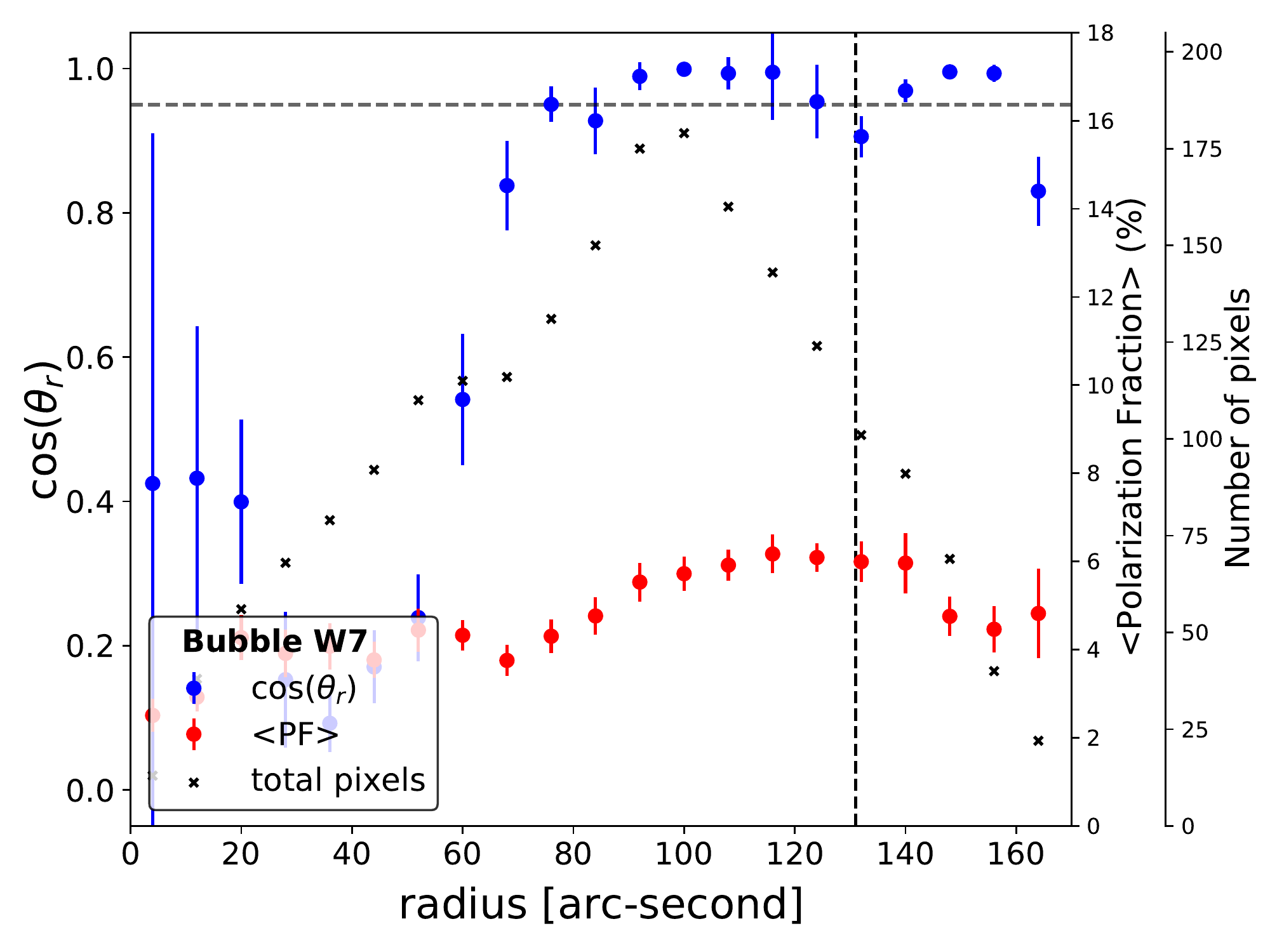}
    \includegraphics[scale=0.3, trim={0cm 0cm 0cm 0cm},clip]{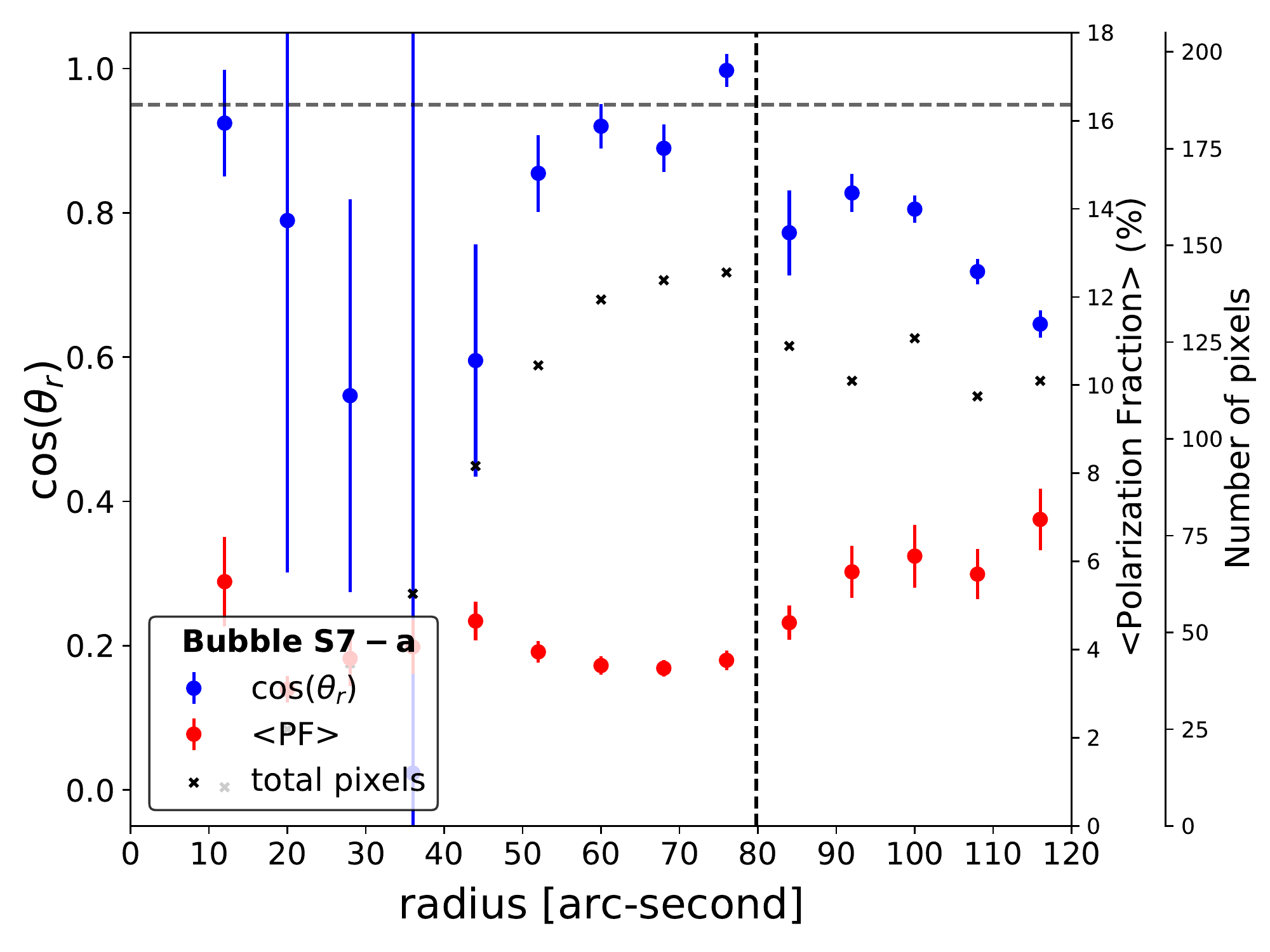}
    \includegraphics[scale=0.3, trim={0cm 0cm 1cm 0cm},clip]{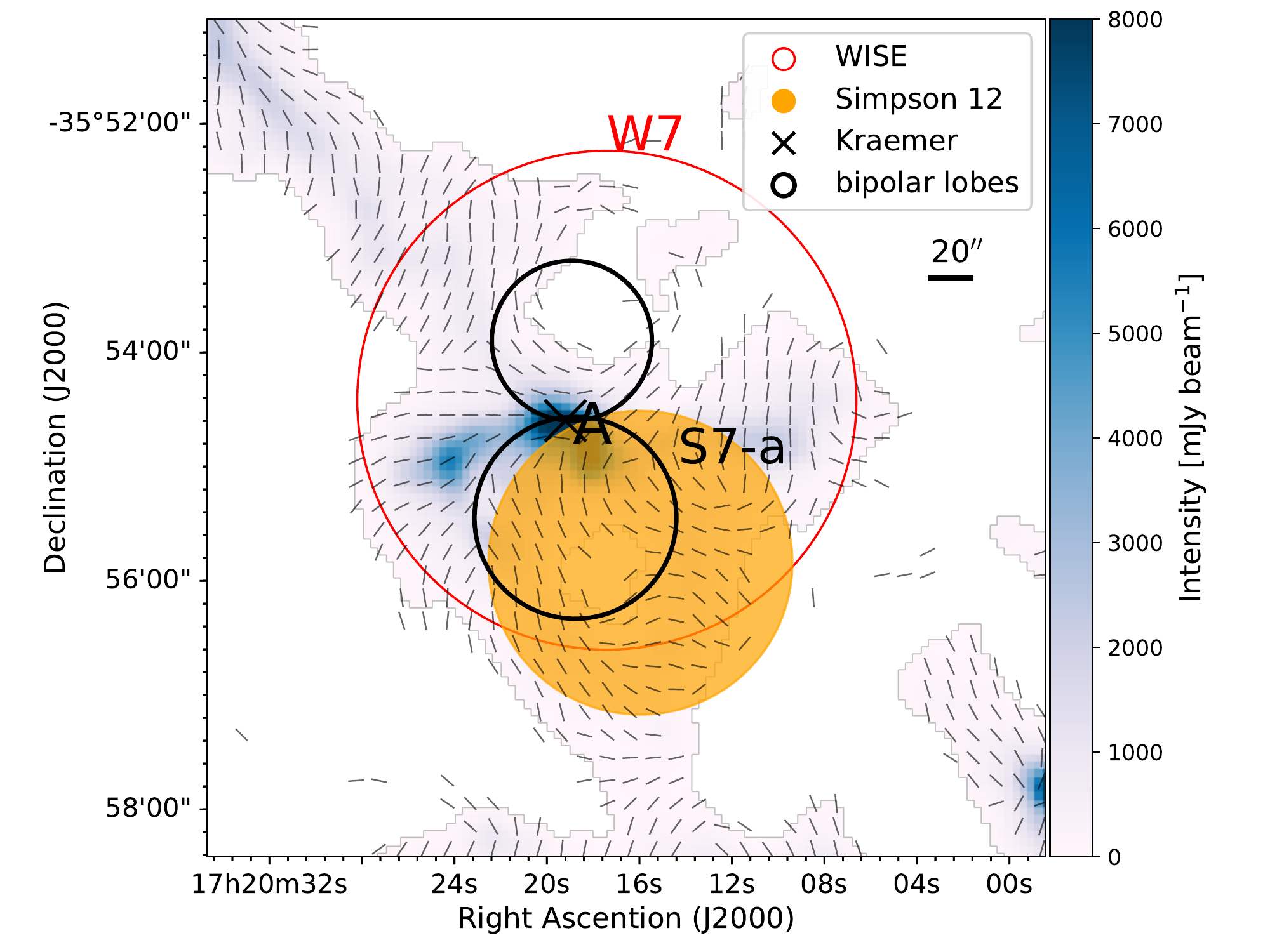}\\
  \end{tabular}%
\caption{Radial polarization in W5 and W7. \textbf{Left column}: The radial profile of $\cos(\theta_r)$ and the mean polarization fraction in each bubble are shown.  The vertical and horizontal dashed lines represent the radius of each bubble and  $\cos(\theta_r) = 0.95$, respectively.  The $\cos(\theta_r)$ values, the mean polarization fraction, and  the total number of pixels (which satisfy the selection criteria of SNR($I$)$ >10$ and SNR($PI$)$> 3$) in each shell are indicated by the blue, red, and black markers, respectively. The polarization fraction error bars represent the standard deviation of the mean (see  Appendix~\ref{apndx:radialPolErrorBar} for $\cos(\theta_r)$ error bars). \textbf{Middle column}: Radial polarization of S5 and S7-a, with the same description as the left column. \textbf{Right column:} Zoomed-in view of each bubble is shown where the magnetic field lines (of identical size) are overlaid on the Stokes $I$ map.  }
\label{fig:RadialPol3column}
\end{figure*}

\begin{figure*}[thbp]
\centering
    \vspace{0.2cm}
    \centering\small (a) Bubble W2, WISE category G.
  \begin{tabular}{c c c}
    \includegraphics[scale=0.3, trim={0cm 0cm 0cm 0cm},clip]{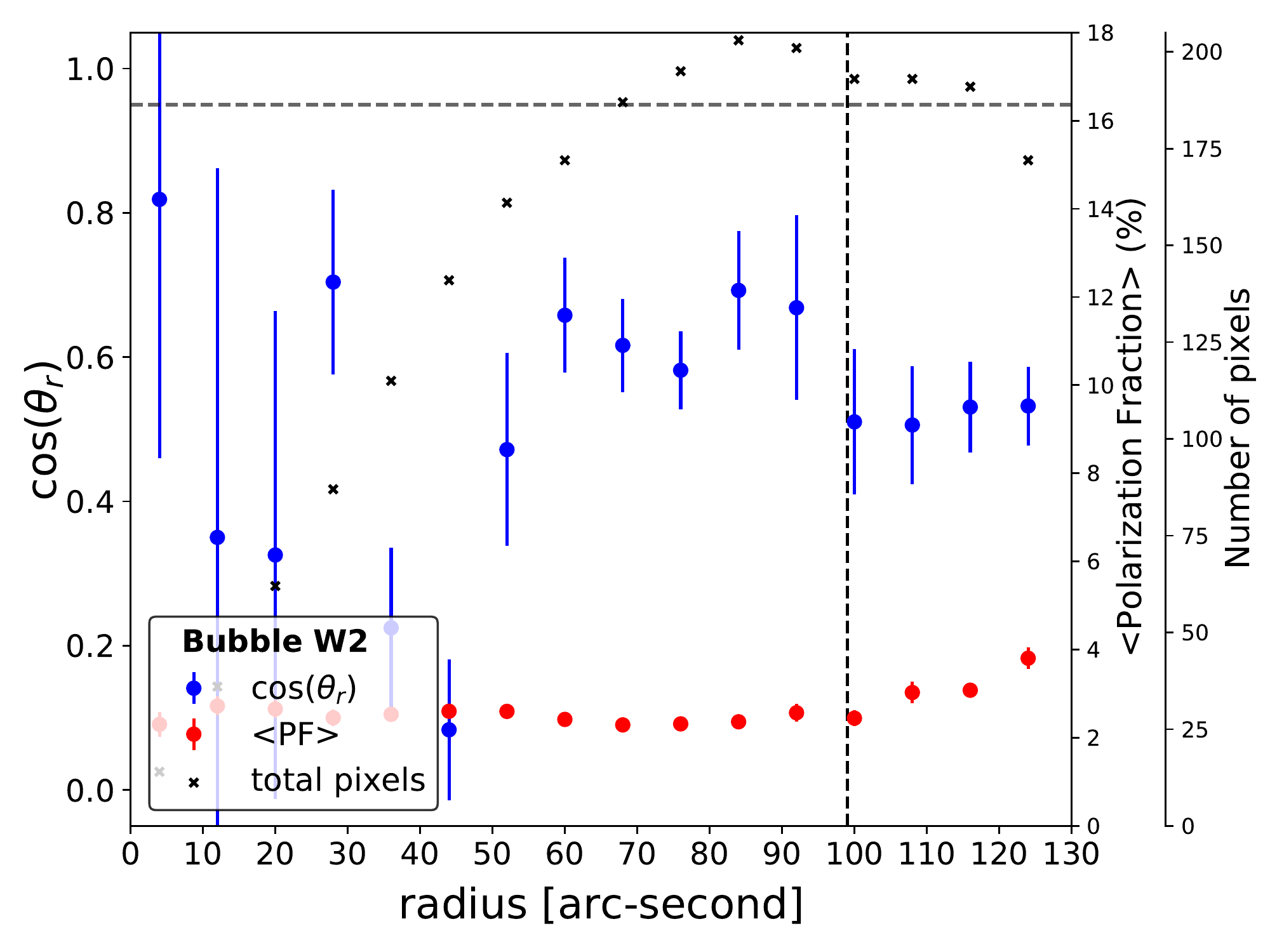}
    \includegraphics[scale=0.3, trim={0cm 0cm 0cm 0cm},clip]{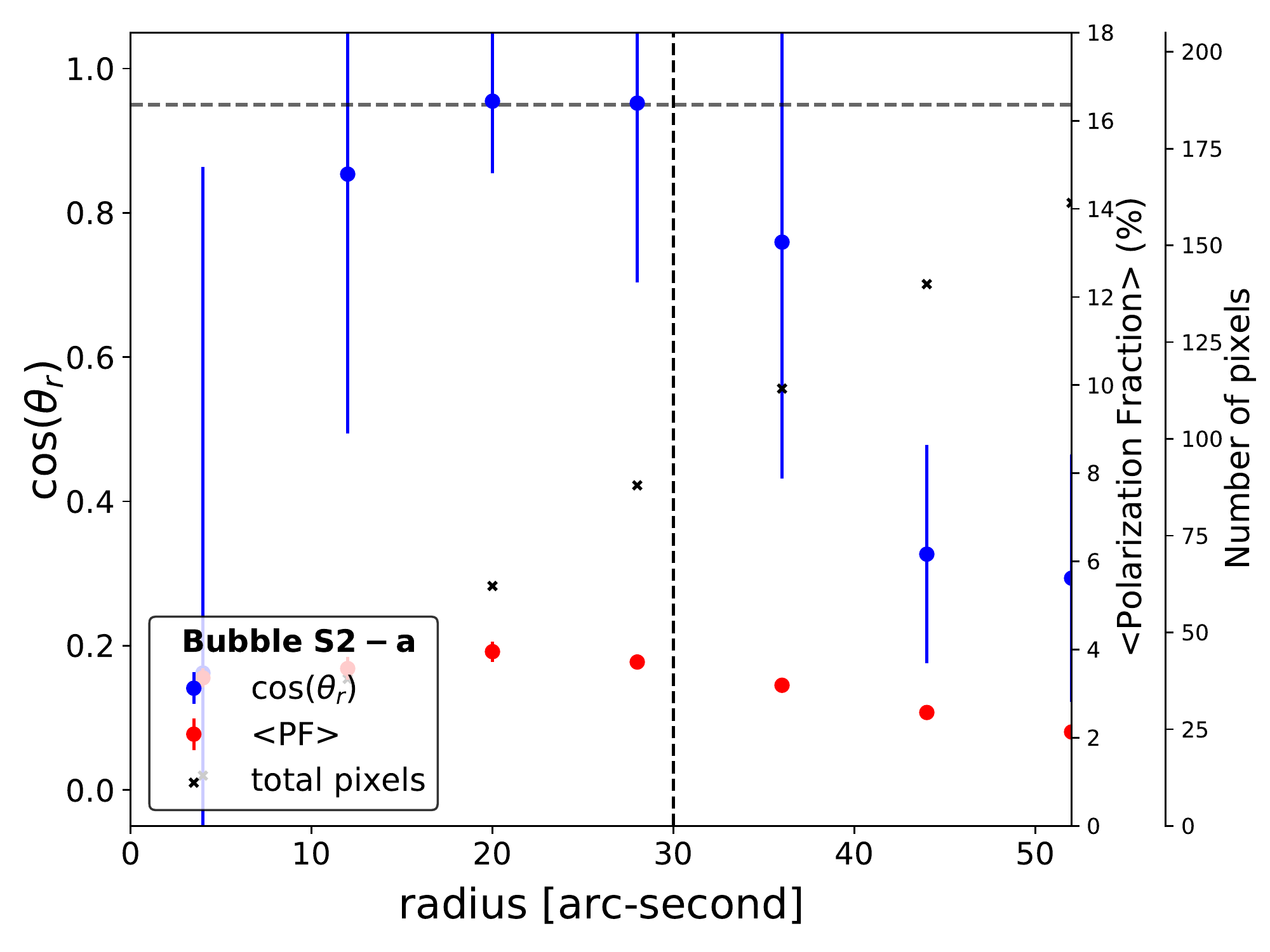}
    \includegraphics[scale=0.3, trim={0cm 0cm 1cm 0cm},clip]{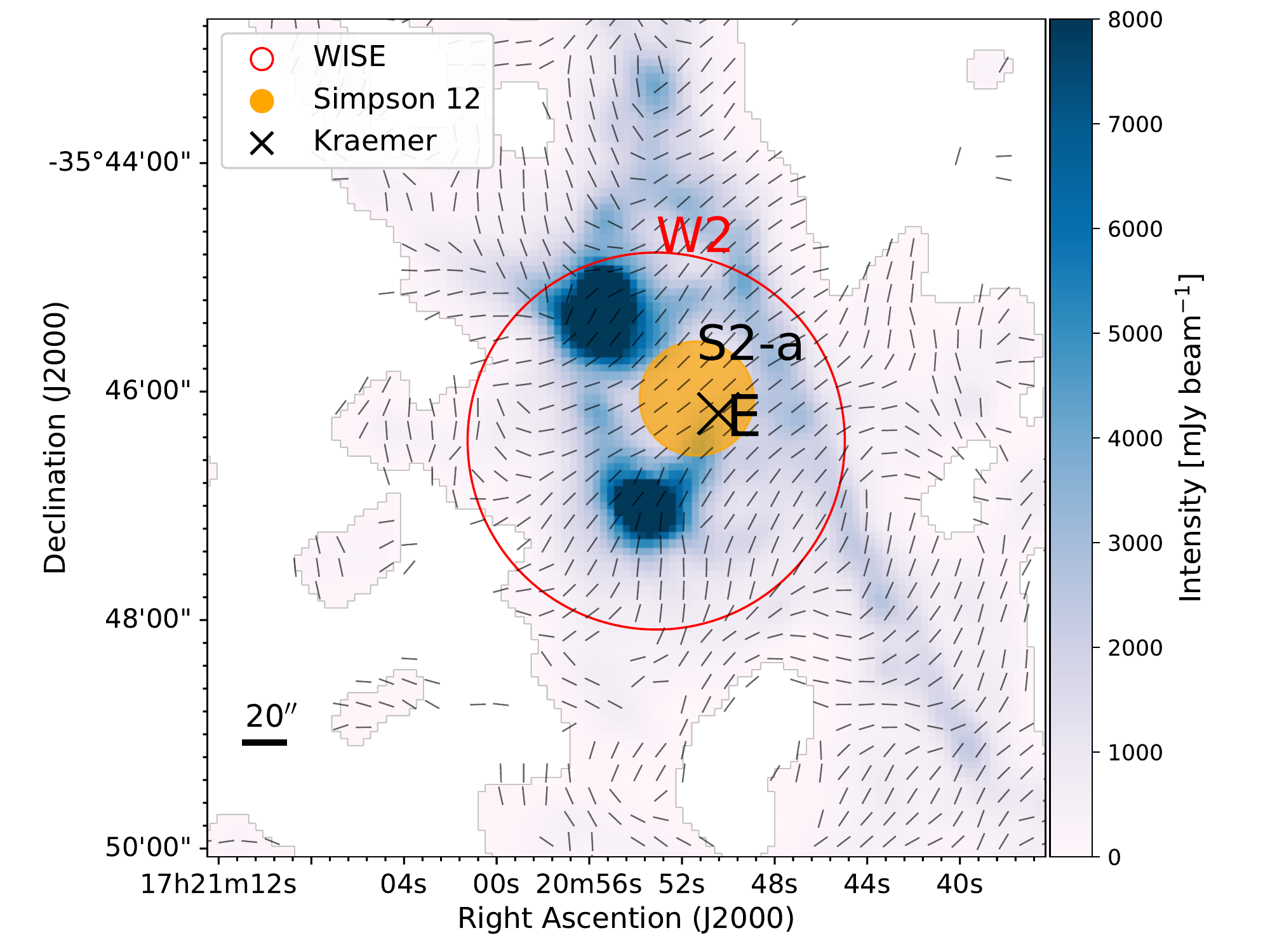}\\
    \vspace{0.05cm}
  \end{tabular}%
   \\ \centering\small (b) Bubble W3, WISE category Q.
  \begin{tabular}{c c c}
    \includegraphics[scale=0.3, trim={0cm 0cm 0cm 0cm},clip]{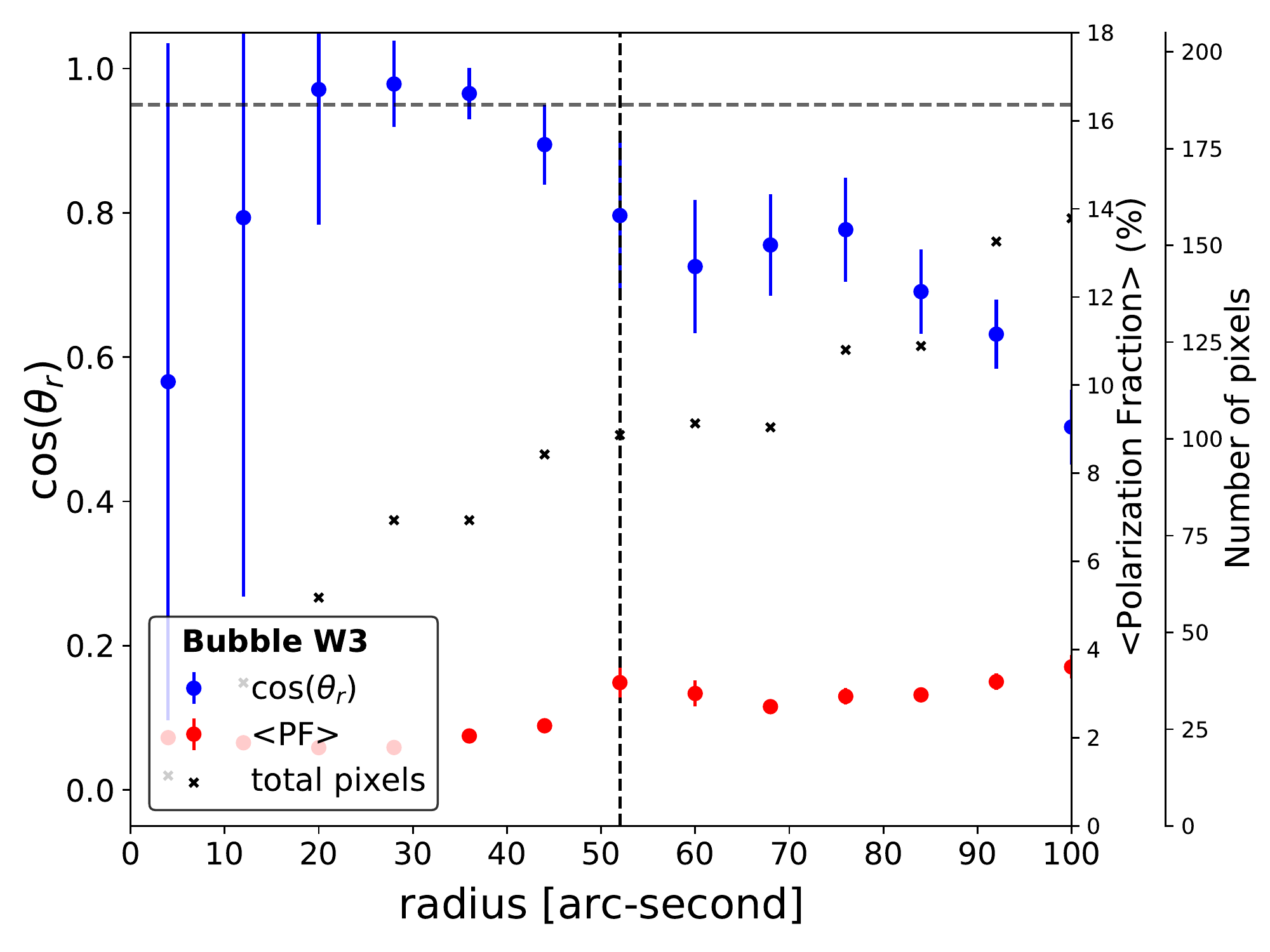}
\includegraphics[scale=0.3, trim={0cm 0cm 0cm 0cm},clip]{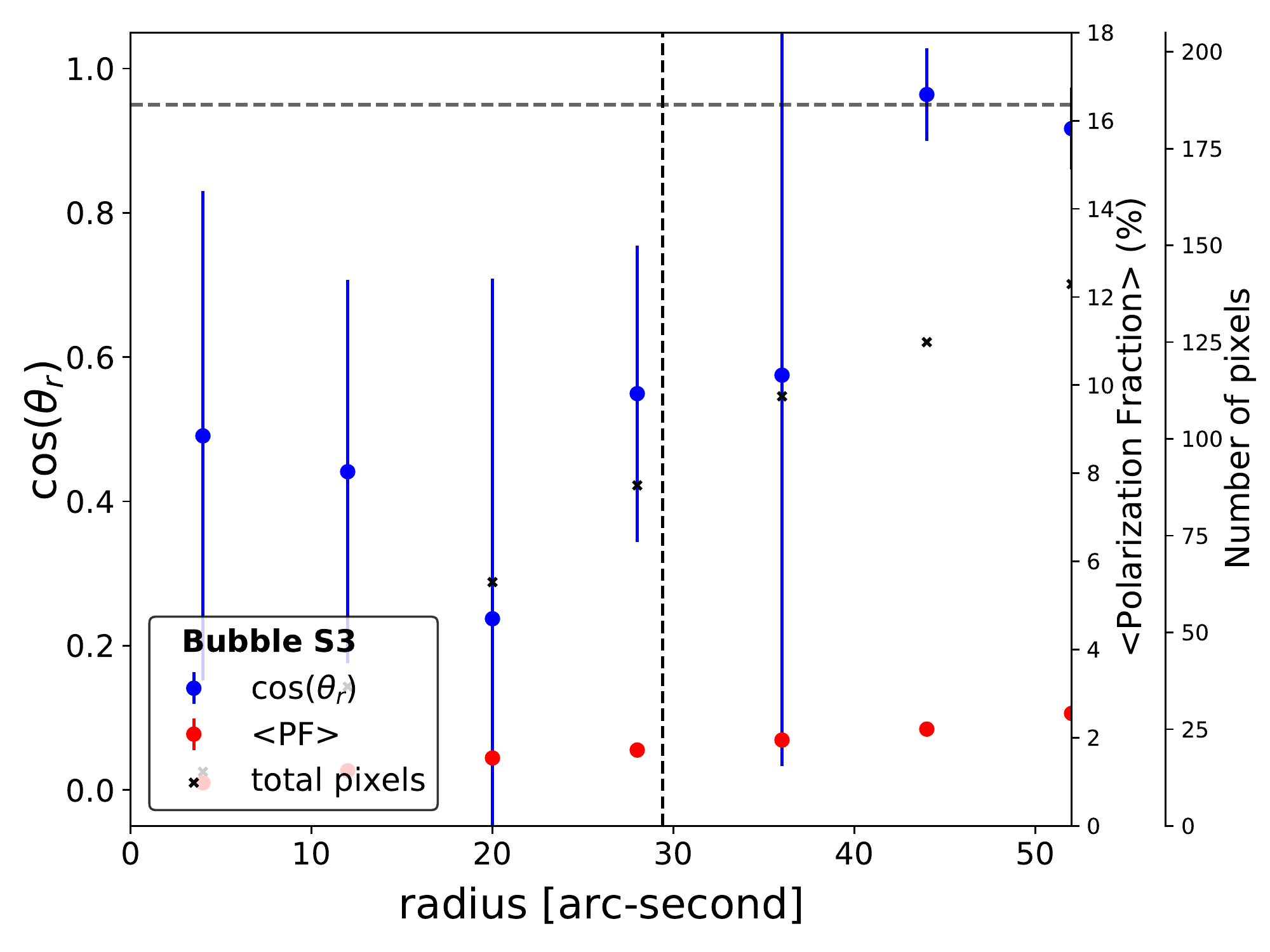}
\includegraphics[scale=0.3, trim={0cm 0cm 1cm 0cm},clip]{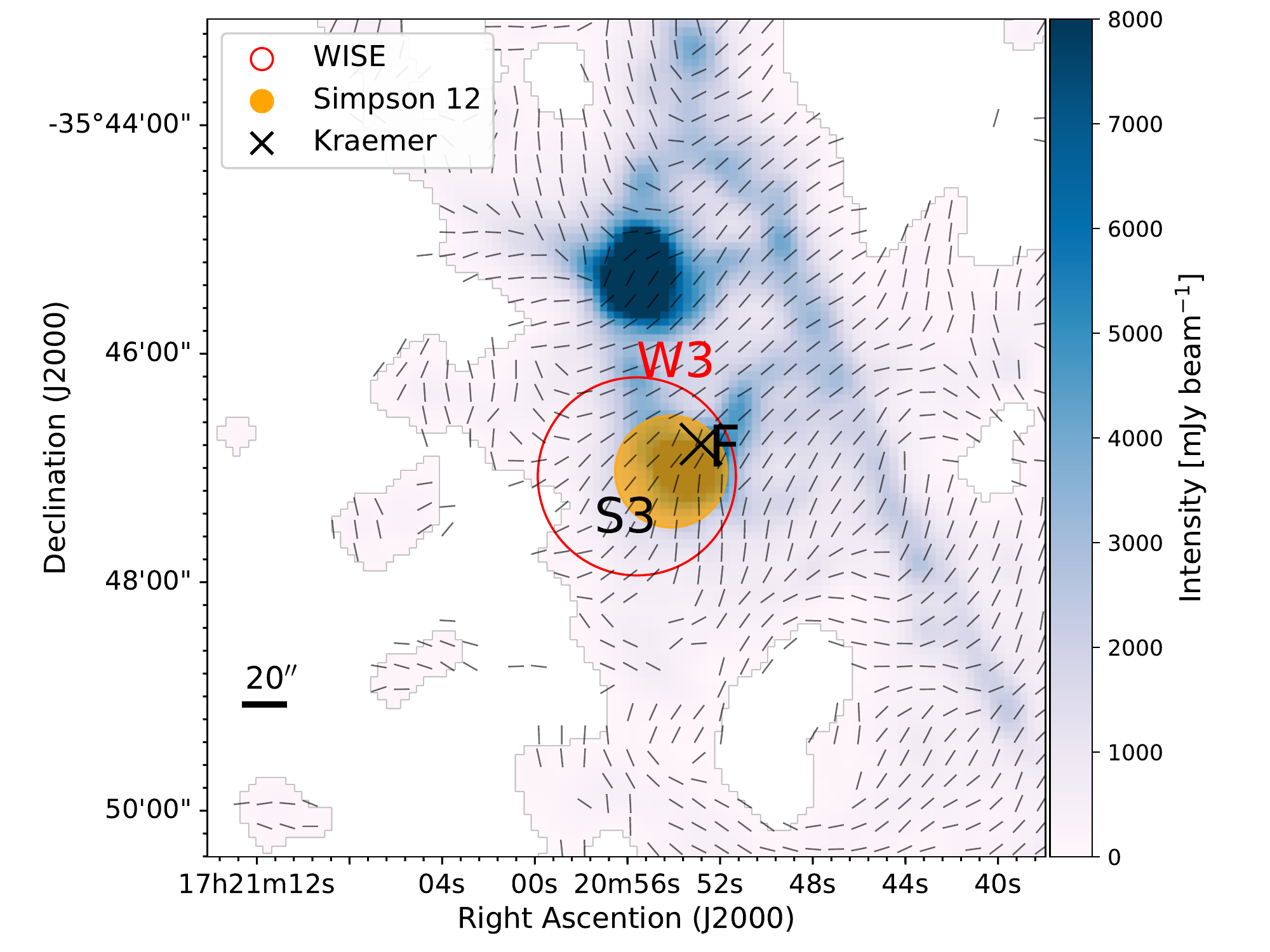}\\
  \end{tabular}%
\caption{Absence of radial polarization in W2 and W3. \textbf{Left column}: Radial profile of $\cos(\theta_r)$ (blue marker) and the mean polarization fraction (red marker) in each bubble are shown.  The radius of each bubble and  $\cos(\theta_r) = 0.95$ are represented by the vertical and horizontal dashed lines, respectively.  In polarization fraction, the error bars  represent the standard deviation of the mean.  Appendix~\ref{apndx:radialPolErrorBar} discusses the error bars for the blue markers. \textbf{Middle column}: Radial polarization of S2-a and S3, with the same description as the left column. \textbf{Right column:} Zoomed-in view of each bubble, representing the observed magnetic field lines (of same size).   }
\label{fig:NoRadialPol}
\end{figure*}

Figures~\ref{fig:RadialPol2column}, \ref{fig:RadialPol3column},  and \ref{fig:NoRadialPol} illustrate the radial profile of both $\cos(\theta_r)$ and the mean polarization fraction for each bubble (averaged in each annulus with a thickness of $8\arcsec$, or two pixels). In these figures, the left panel illustrates $\cos(\theta_r)$ and the mean polarization fraction with blue and red marks, respectively. The right panel shows a zoomed-in view of the magnetic field lines of each bubble overlaid on Stokes $I$. The dashed vertical and horizontal lines indicate the radius of each bubble (the Radius and R$_{\text{eff}}$ columns in Tables~\ref{table:Wise} and \ref{table:Simpson}, respectively) and  $\cos(\theta_r) = 0.95$, respectively. The blue markers indicate the total number of pixels (that satisfy the selection criteria of SNR($I$)$ >10$ and SNR($PI$)$> 3$) in each annulus. Some bubbles have boundaries outside of the NGC~6334 cloud in regions with the lowest density and fewest pixels.

We find that radial polarization in bubbles is accompanied by  a higher polarization fraction, implying that \HII\ regions have ordered and compressed the magnetic field lines. A statistical study involving a large number of bubble polarization observations is necessary to determine if the amount of increase in polarization fraction can quantitatively reveal information about the physical properties of the bubbles (e.g., ratio of regular to random field or Compression factor).

To facilitate discussion of these findings, we divide this section into two subsections: bubbles that exhibit radial polarization at one or more radial distances and bubbles that do not. Figures~\ref{fig:RadialPol2column} and \ref{fig:RadialPol3column} illustrate bubbles with radial polarization at one or more radial distances. When a bubble is associated with a Simpson bubble,  a middle panel is added to display the results for the Simpson bubble (Figure~\ref{fig:RadialPol3column}). Figure~\ref{fig:NoRadialPol} illustrates bubbles with no radial polarization patterns at any radial distance.

\subsubsection{Bubbles with radial polarization signatures}
\label{sec:radialBubbles}

Radial polarization at one or more radial distances can be seen in bubbles W1 and W5 to W8, with W5 and W7 having corresponding Simpson bubbles (S5 and S7-2). W1, W6, and W8 (with no corresponding Simpson bubbles) are illustrated in Figure~\ref{fig:RadialPol2column}, while W5 and W7 are depicted in  Figure~\ref{fig:RadialPol3column}.

{\textbf{Bubble~W1}} is classified as a Radio Quiet \HII\ region in the WISE catalog, with an unidentified distance by the WISE catalog or \citet[][]{Wuetal2014}. However, its dust morphology indicates that W1 is associated with NGC~6334.  When the location of W1  
is compared with the morphology of sub-millimeter dust observations shown in Figure~\ref{fig:NGC6334Bubbles}, it appears as though the center of this bubble, as identified in the WISE catalog, is slightly shifted relative to the center location where the effects of bubble expansion on dust can be observed. 
At $45\arcsec$ radial distance, the polarization exhibits a radial pattern, with a peak in polarization fraction (around 7\%), as illustrated in the upper-left panel of Figure~\ref{fig:RadialPol2column}. Following that at greater distances, polarization lines diverge from radial angle and the polarization fraction decreases.

{\textbf{Bubble~W6}} is categorized as a Known bubble, and it is approximately the same distance away from us as NGC~6334. When projected onto the plane of the sky, W6 appears to contain a $\sim 1$\,pc-long filament. The middle-left panel of Figure~\ref{fig:RadialPol2column} indicates tangential polarization near the center of the bubble (along the filament) and radial polarization between radii of $\sim 60\arcsec$ to $110\arcsec$. This radial polarization range also exhibits higher polarization fractions ($6-7\%)$. As depicted in Figure~\ref{fig:NGC6334Bubbles}, W6 appears in contact with W5 to the northwest and W7 to the southeast (when projected onto the plane of the sky). Therefore, at greater distances beyond the radius of W6 the effects of radial polarization associated with W5 and W7 (higher polarization fractions) can be seen. 

Although {\textbf{Bubble~W8}} is classified as a Known bubble, neither the WISE catalog nor the study by \citet[][]{Wuetal2014} indicate its distance. However, the dust morphology suggests that W8 should be directly associated with NGC~6334. The lower-left panel of Figure~\ref{fig:RadialPol2column} indicates radial polarization in the radius range of $160\arcsec$ to $180\arcsec$, along with locally maximal polarization fraction ($\sim 9\%$). 

{\textbf{Bubble~W5}}, a Known bubble, has the same distance from us as NGC~6334. The magnetic field lines associated with this bubble clearly demonstrate the impact of the \HII\ region, as illustrated in the inset of  Figure~\ref{fig:NGC6334Bubbles}. The upper-left panel of Figure~\ref{fig:RadialPol3column} shows radial polarization in the radius range of $\sim 60\arcsec$ to $90\arcsec$ with on-average higher polarization fraction ($\sim 6-8\%$). This is similar to the values in the upper-middle panel of Figure~\ref{fig:RadialPol3column} for S5 (radii of $\sim 50\arcsec$ to $85\arcsec$ and $PF$ of $\sim 6-10\%$),  where the small difference is due to the slight difference in radius and center for W5 and S5. The observed field lines appear to be more  consistent with S5 than W5. At radial distances greater than the radius of the bubble, the polarization pattern becomes non-radial and the polarization fraction decreases by a factor of $\sim 2$, indicating that the field lines are less-ordered. 

{\textbf{Bubble~W7}} is at the same distance as NGC~6334 and in the Known category. The presence of radial polarization and an increase in the polarization fraction can be seen in this bubble  at $r \simeq 90\arcsec - 100\arcsec$. Additionally, at distances greater than the radius of W7, the polarization lines exhibit a higher polarization fraction or radial polarization due to the influences of W6 and W8.

W7 has a bipolar morphology~\citep[suggested by][based on the dust temperature studies of the two lobes]{HarveyGatley1983} and on its south side is associated with S7-a, which is part of a Herbig-Haro object~\citep[southern lobe;][]{Bohigas1992}. This bipolar morphology is likely caused by the confinement of an \HII\ region within a flattened-like (or torus-like) molecular gas structure \citep[][] {KraemerandJackson1999}, which allows the ionized gas to escape only in the north and south directions~\citep[extending $2\arcmin$;][]{Rodriguezetal1988}. This bipolar structure is also visible in our observations; the magnetic field morphology of W7 resembles a radial polarization pattern for two lobes (to the north and south of  its center), with their radial polarization features coinciding near the center of W7, as illustrated in the lower-right panel of Figure~\ref{fig:RadialPol3column}. 

The southern lobe of W7 can be studied using the S7-a bubble. S7-a exhibits a radial polarization pattern at $r \simeq 80\arcsec$, as illustrated in the lower-middle panel of Figure~\ref{fig:RadialPol3column}. However, the polarization fraction increases at radii greater than this value, as the polarization fraction is also influenced by W8 at these radii.

\begin{figure}[thbp]
\centering
\includegraphics[scale=0.4, trim={0cm 0.25cm 0cm 0cm},clip]{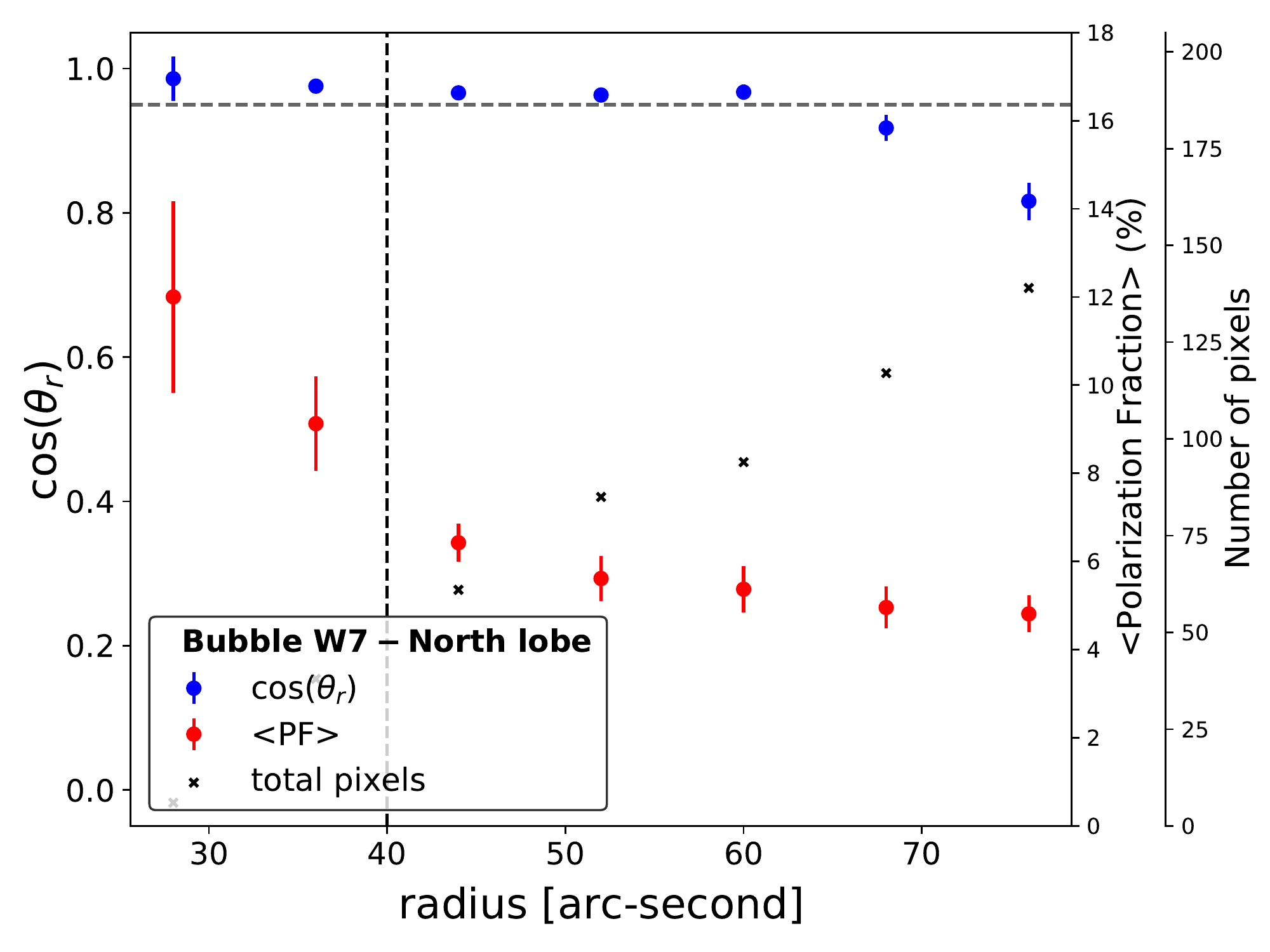}
\caption{Radial polarization study of W7. The mean polarization fraction, the total number of pixels in each annulus, and $\cos(\theta_r)$ are all shown by the red, black, and blue markers.  }
\label{fig:SourceA}
\end{figure}

No distinct \HII\ region has been identified in the WISE or Simpson catalogs  as being directly associated with the northern lobe of W7. We approximate the bipolar structure (shown with black circles in Figure~\ref{fig:SourceA}) based on the 6\,cm observations of \citet[][]{Rodriguezetal1988}, in which the northern lobe appears slightly shifted to the south in comparison to the tangential field lines (radial polarization) visible in our data. As a result, we perform our radial polarization analysis on a bubble with a central location of $(\alpha, \delta) = (260.07^{\circ}, -35.89^{\circ})$ and a radius of $\sim 40\arcsec$, as estimated by the location of our observed polarization lines.  The absence of a distinct \HII\ region for the northern lobe in the catalogs emphasizes the importance of polarization studies in identifying and examining molecular cloud substructures. We discuss this in more detail in Section~\ref{sec:CHT}, where we investigate the possibility of using polarization properties to identify bubbles that have impacted the field lines.

\subsubsection{Bubbles without radial polarization}

{\textbf{Bubble~W2}} is classified as Group in the WISE catalog, with an undetermined distance (to our knowledge). As illustrated in the upper-left panel of  Figure~\ref{fig:NoRadialPol}, the bubble has polarization angles of $\sim 55^{\circ} (\cos{\theta_r} \simeq 0.6)$ relative to radial direction and is therefore closer to tangential than to radial lines. These fields are oriented in a single direction (perpendicular to the main ridge) and are likely dominated by the region's high density cores and/or clumps. 
This bubble demonstrates a nearly constant polarization fraction of $\sim 2.5\%$. Similarly, compared to other bubbles the averaged polarization fraction is lower in S2-a (maximum 4\% as illustrated in the upper-middle panel of Figure~\ref{fig:NoRadialPol}), which is nested within W2. We note a small increase in polarization fraction just behind the radius of S2-a, followed by a slight decrease beyond the radius. While the polarization lines appear to have an average $\cos{\theta_r} \simeq 0.95$ at $20\arcsec$ and $30\arcsec$, their large error bars indicate a non-radial polarization pattern. The large error bars for $\cos{\theta_r}$ are due to the fact that the field lines are mostly in one orientation, and thus  have different radial angles (relative to radial direction) at various  points on the circle. 

{\textbf{Bubble~W3}} is classified as a radio Quiet bubble and appears within W2, with an undetermined distance in the WISE catalog.  W3 and its corresponding Simpson bubble, S3, do not indicate radial polarization, as illustrated in the lower-left and lower-middle panels of Figure~\ref{fig:NoRadialPol}. At $r \simeq 35\arcsec$, it appears as though the polarization in W3 is approaching radial. However, because the error bars fall below $\cos(\theta_r) =0.95$, this bubble {\textit{does not}} pass our criteria for radial polarization and further analysis is required. As with W2, the polarization lines in W3 appear to be unaffected by the \HII\ region, with a nearly constant polarization fraction ($\sim 2$\%) inside the bubble and a slight increase near the radius of the bubble (up to $\sim 3\%$). 
W3 (and S3) belong to a dense region of NGC~6334 with active star formation.

While W2 and W3 have unidentified distances in the WISE catalog and do not exhibit radial polarization, observations indicate that they (or their corresponding Simpson bubbles) are associated with NGC~6334. S3 and S2-a  are both compact \HII\ regions in the dense part of NGC~6334. A cluster of B0-B$0.5$ Zero Age Main Sequence stars likely ionize the S2 region~\citep{Tapiaetal1996}. S3 is associated with the strongest maser in NGC~6334, H$_2$O, OH, and CH$_3$OH masers~\citep{MoranRodiguez1980, Batrlaetal1987, ForsterCaswell1989,MentenBatrla1989, KraemerandJackson1999}, and a young molecular outflow~\citep{BachilerCernicharo1990}. 

Finally, we note that as illustrated in Figure~\ref{fig:NoRadialPol}, bubbles with no evidence of radial polarization have larger error bars (indicating a greater dispersion of $Q_r$ and $U_r$ values within each annulus). As the angle changes with respect to the radius of the bubble, a uniform magnetic field morphology results in greater dispersion, whereas tangential field lines result in  small dispersion and error bars.

\section{Discussion}
\label{sec:discussion}
All bubbles, except for W2 (and S2-a) and W3 (and S3) exhibit some degree of radial polarization. Table~\ref{table:ImpactedRange} summarizes our findings regarding the magnetic field morphology of \HII\ regions within NGC~6334 as discussed in Section~\ref{sec:radial}. Examining the energy balance associated with these regions is required to better understand magnetic field geometries. Additionally, the polarization patterns of the cloud  from small to large scales can reveal important information about substructures within the cloud and the cloud's evolution and/or formation. We discuss these in the following subsections.  

\subsection{Field strengths and energy comparisons}

\begin{table*}[t]
\centering
\hspace{-1cm}
 \begin{tabular}{| c  c  c  c  c c|} 
 \hline
Bubble & radius ($\arcsec$) &  source$^{\gamma}$ & $B_{\textsc{LOS}}^{\dagger}$ ($\mu$G)& $\langle B_{\textsc{POS}}\rangle ^{\star}$ ($\mu$G) & RP evidence    \\\hline\hline
W1	   & 68.0 &  & & 42 &   at $45\arcsec$               \\
W2	   & 99.0 & E & & 360 &   --          \\
S2-a   & 30 & E  &\hspace{-1.2cm}\makecell[l]{$-263\pm78$ at 1665\,MHz (OH)\\$-340\pm78$ at 1667\,MHz (OH)\\$-169\pm33$ at 1420\,MHz (\HI )}  &  --  &   --                    \\
W3	   & 52.0 & F & & 415 &   --          \\
S3     & 29.4 & F &  & 351   &   --                    \\
W5	   & 107.0 & D &\hspace{-1.2cm}\makecell[l]{$-60\pm46$ at 1665\,MHz (OH) \\ $-69\pm58$ at 1667\,MHz (OH)\\ $-93\pm13$ at 1420\,MHz (\HI ) } &  99   &   $60\arcsec$ to $90\arcsec$          \\
S5    & 86.4  &  D & & 92 &   $50\arcsec$ to $85\arcsec$          \\
W6	  & 119.0 & C &  & 144  &   ~$60\arcsec$ up to $110\arcsec$     \\
W7	  & 131.0 & A & \hspace{-1.2cm}\makecell[l]{$+148\pm20$ at 1665\,MHz (OH) \\ $+162\pm33$ at 1667\,MHz (OH)\\ $+47\pm15$ at 1420\,MHz (\HI )}  & 132 &   $85\arcsec$ to $110\arcsec$  \\
S7-a  & 79.8 & A & & 114  &   at $75\arcsec$  \\
W8	  & 171.0 & \RN{5} & & --  &   at $160\arcsec$ and $160\arcsec$ to $180\arcsec$  \\
\hline
\end{tabular}
\caption{Summary of radial polarization detection in bubbles. The first, second, and fourth columns show bubble numbers, bubble radius, and  magnetic field strength obtained from literature, respectively.  The $\gamma$ represents sources identified by \citet[][]{Rodriguezetal1982} that were used by \citet[][]{Sarmaetal2000} and \citet[][]{Balseretal2016} for Zeeman measurements. The $\dagger$  denotes VLA Zeeman measurements by \citet{Sarmaetal2000}, with the negative and positive signs indicating directions toward and away from us, respectively.  We note that the opposite sign convention of Zeeman measurements is used in Faraday measurements. 
The fifth column denoted by $\star$ indicates the approximate plane-of-sky magnetic field strength in the molecular gas surrounding the bubble, using strengths obtained by Paper~\RN{1}.  The values listed are the weighted means of the magnetic field strengths estimated by Paper~\RN{1} in filaments overlapping with the bubbles. The final column displays the regions within each bubble that exhibit radial polarization. \citet[][]{Balseretal2016} determine magnetic strength of source A to be $190 \pm 96\,\mu$G and the strength of source D to be approximately 180 to 1200\,$\mu$G for Source D. }
\label{table:ImpactedRange}
\end{table*}

We employ established equations and observations from the literature to determine the energetic significance of the \HII\ regions in comparison to magnetic fields. These are discussed in greater detail in the following two subsections.

\subsubsection{Energy balance relations}
\label{sec:Energy}
To estimate the gas thermal pressure ($P_{\rm{gas}}$),  radiation pressure ($P_{rad}$), dynamic pressure ~\citep[$P_{dyn}; $ e.g.,][due to bubble expansion]{Paveletal2012}, magnetic  pressure~\citep[$P_{B}$; e.g.,][]{Pattleetal2022PP7}, and magnetic tension~\citep[$T_{B}$; e.g.,][]{BoularesCox1990} per unit area (due to curved field lines) of bubbles, we use the following equations (for tangential field lines):
\begin{equation}
\begin{aligned}
&P_{\rm{gas}} = n k_B T,\\
&P_{rad} =\frac{L}{4\pi r^2 c},\\
&P_{dyn} = 0.5\rho v_{ex}^2,\\
&P_{B} = \frac{B^2}{8\pi},\\
&T_{B} = \frac{B^2}{4\pi},
\label{eq:pressures}
\end{aligned}
\end{equation} 
where $n = n_e$, $k_B$, $T$, $c$, $r$, $L$, $\rho$, $v_{ex}$, and $B$  are the electron volume density, Boltzmann constant, gas temperature, speed of light, distance to the center of the bubble, luminosity, mass density, expansion velocity, and magnetic field strength (perpendicular to gas motion), respectively.

While the above-mentioned $P_{rad}$ is used for optically thick regions, studies~\citep[][]{Reissletal2018Rad} show that  radiation pressure is generally small (or negligible) compared to gravity or gas pressure. We can estimate the magnetic field strength ($B_{\rm{resist}}$) required to resist the combined effects of gas, radiation, and dynamic  pressures, using:
\begin{equation}
P_{\rm{B, resist}} + T_{\rm{B, resist}} = \frac{3 B_{\rm{resist}}^2}{8\pi} = P_{gas} + P_{rad} + P_{dyn}.
\end{equation}
Magnetic fields significantly less than this $B_{\rm{resist}}$ value indicate that the field lines are altered by the \HII\ regions. 

\subsubsection{Parameter estimates}
\label{sec:parameters}
\begin{figure}[thbp]
\centering
\includegraphics[scale=0.5,clip, trim={1cm 0cm 2.5cm 1.5cm}]{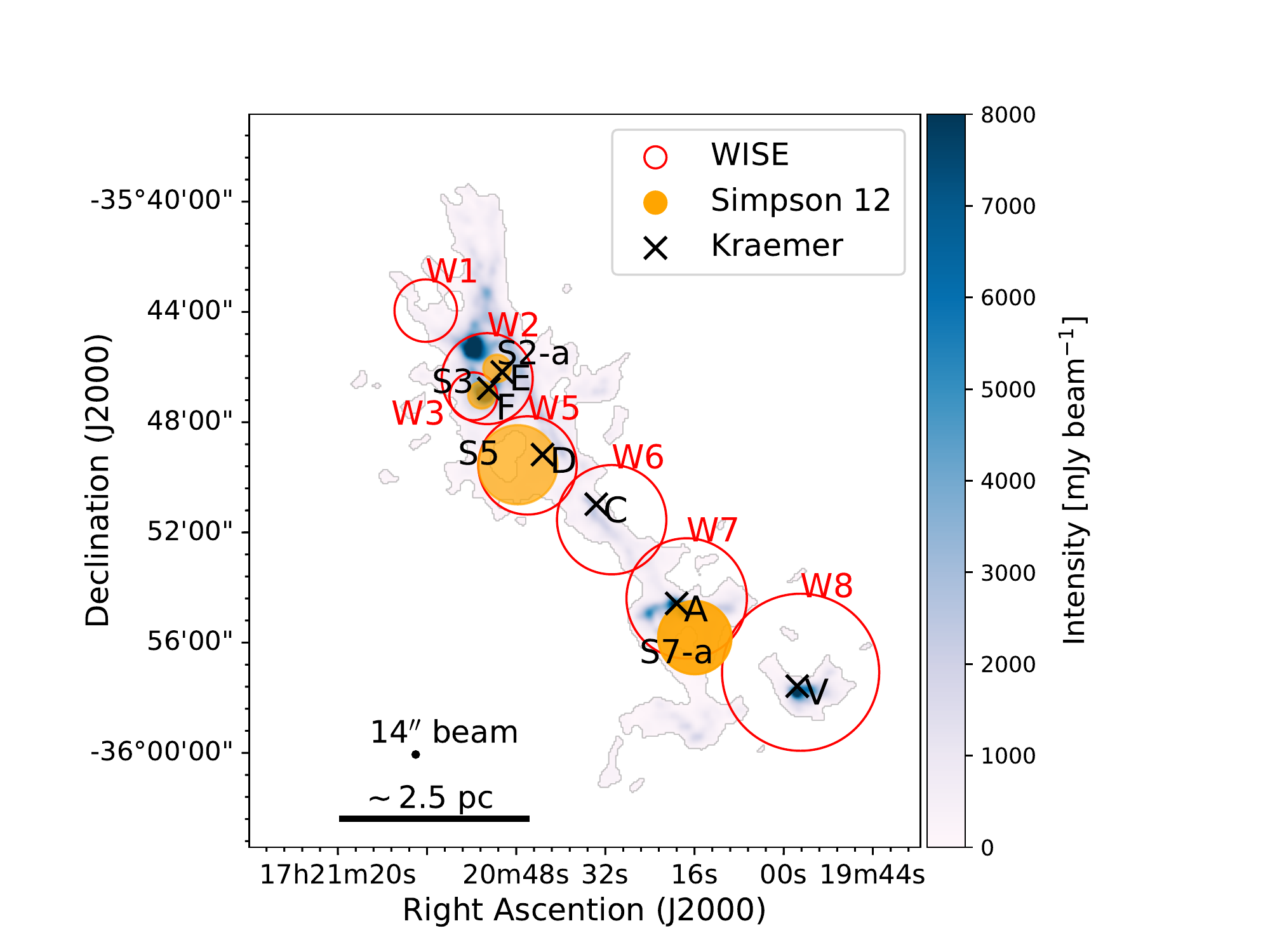}
\caption{Sources identified by \citet{KraemerandJackson1999}. The background color image shows the Stokes $I$ map. The orange and red circles show the \HII\ regions identified in the WISE and \citet{Simpsonetal2012} catalogs, respectively. The x marks show the sources in \citet{KraemerandJackson1999} and used in \citet{Sarmaetal2000} to perform Zeeman measurements. }
\label{fig:kramerSources}
\end{figure}

Except for W1, all of these bubbles have been previously identified in various observations~\citep[e.g.,][sources denoted as A to E and source \RN{5} shown in Figure~\ref{fig:kramerSources}]{Rodriguezetal1982} and their magnetic fields were studied using Zeeman measurements~\citep[][see Table~\ref{table:ImpactedRange}]{Sarmaetal2000, Balseretal2016}. We note that \citet{KraemerandJackson1999} found an anti-correlation between the presence of dense gas and the 6\,cm radio flux, which they attribute to gas dispersion by the feedback from the hottest stars.  This gas and dust dispersion is also visible in our $850\,\mu$m dust observations. 

We also employ the parameters listed in Table~\ref{table:KramerTable} obtained from \citet{Rodriguezetal1982} and \citet[][]{Russeiletal2016}. \citet{Rodriguezetal1982} assumed a temperature of 10$^{4}$\,K  and found the electron volume densities and the \HII\ masses of these sources. \citet[][]{Russeiletal2016} provide the expansion velocities ($v_{ex}$; see their Table~1). 
Using the $L$ and $n_e$ values from \citet{Rodriguezetal1982}, a temperature of $10^4$\,K, and the $v_{ex}$ values, we determine the $B_{\rm{resist}}$ magnetic field strengths as listed in Table~\ref{table:KramerTable} for sources A, C, D, E, and F.  
These values (except for sources F and E) exceed the  observed magnetic field strengths (see Table~\ref{table:ImpactedRange}) determined using Zeeman observations~\citep[][]{Sarmaetal2000, Balseretal2016} and the Davis-Chandrasekhar-Fermi technique~\citep[DCF;][associated with substructures on the shells]{DavisGreenstein1951, ChandrasekharFermi1953}  described in Paper~\RN{1}. 

\renewcommand{\arraystretch}{1.4}
\begin{table*}[t]
\hspace{-1.5cm}
\centering 
 \begin{tabular}{| l  l  l  l l l l l l l  |} 
 \hline
Src &  \shortstack{$\alpha$ \\($^{\circ}$J2000)} & \shortstack{$\delta$ \\ ($^{\circ}$J2000)} &  \shortstack{Radius \\ ($\arcsec$)} &  \shortstack{T$_k$ \\(K)}  & \shortstack{$n_e$ \\($\times 10^3$\,cm${^{-3}}$)} & \shortstack{M$_{\rm{H\protect\scaleto{$II$}{1.2ex}}}$\\ (M$_{\odot}$)} & \shortstack{$L$\\ ($L_{\odot}$)} & \shortstack{$v_{ex}$ \\(km\,s$^{-1}$)} & \shortstack{$B_{\rm{resist}}$ \\ $\mu$G}\\ \hline \hline
A   &   260.08    &  $-35.91$     &  100    &  60     & 20    & 0.4   &   $8\times 10^4$ & 15.7 &  534      \\
C   &   260.14    &  $-35.85$     &  40     &  70-80  & 2     & 3     &   $7\times 10^4$ & 14   &  267       \\
D   &   260.18    &  $-35.82$     &  40     &  40-50  & 3     & 5     &   $2\times 10^5$ & 14   &  308       \\
E   &   260.21    &  $-35.77$     &  20     &  50-60  & 7     & 1     &   $8\times 10^4$ & 15$^{\bullet}$   &  397       \\
F   &   260.22    &  $-35.78$     &  10     &  60     & 40    & 0.03  &   $3\times 10^4$ & 15$^{\bullet}$   &  744      \\
V   &   259.99    &  $-35.96$     &  30     &  50-60  &       &       &                  &         &                   \\
\hline
\end{tabular}
\caption{Sources studied in \citet{KraemerandJackson1999}, \citet{Rodriguezetal1982}, and \citet[][]{Balseretal2016} as shown in Figure~\ref{fig:kramerSources}. The kinetic temperature, radius, volume and column density are listed for each source from \citet{KraemerandJackson1999} and the electron volume density, \HII\ mass, and stellar luminosity are from \citet{Rodriguezetal1982}. The  expansion velocities from~\citet[][see their Table~1]{Russeiletal2016} are denoted by $v_{ex}$. The $v_{ex}$ values denoted by $\bullet$ are estimates based on the available $v_{ex}$ of other bubbles.  $B_{\rm{resist}}$ is the minimum magnetic field strength needed to resist the bubble's impact. These sources are associated with the WISE and Simpson bubbles. }
\label{table:KramerTable}
\end{table*}

\subsubsection{Energy balance of individual bubbles: Bubbles with radial polarization signatures}

W1, W5 (and S5), W6, W7 (and S7-a), and W8 all exhibit radial polarization. Among these, W5 (and S5) and W7 (and S7-a) have Zeeman observations associated with them. 
\textbf{W5 and S5} correspond to source D, which is identified as an extended, amorphous, and roughly spherical region. Source D contains H$_2$O maser observations ~\citep[][]{MoranRodiguez1980} and has a lower  molecular hydrogen density than any other source in NGC~6334~\citep[][]{KraemerandJackson1999}. We found that a magnetic field strength of 308\,$\mu$G is required to resist the bubble, which is approximately a factor of three greater than the magnetic fields observed with Zeeman or DCF in the vicinity of this region. Therefore, we suggest that the field lines in this region are  pushed and influenced by the gas pressure as evident by the radial polarization (tangential magnetic field) morphology. 

The center of \textbf{W6} coincides with source C. \citet{Sarmaetal2000} found no OH absorption toward the continuum peak of this source to detect Zeeman splitting. This source is non-spherical and likely contains an O7 star~\citep{Strawetal1989, KraemerandJackson1999}. When projected onto the plane of the sky, W6 appears to contain a dense filament, as illustrated in the middle-right panel of Figure~\ref{fig:RadialPol2column}. This filament can be located inside, at the front, or at the back of the bubble in three dimensions.  Magnetic fields run parallel to this filament on the plane of the sky, while the edges of the bubble  exhibit radial polarization behavior. 

Three explanations can be made for the magnetic morphology of the filament associated with W6: 1) The dense filament may be located on the shell (in 3D; foreground or background of the bubble) and thus the field lines may appear running parallel to the filament due to their tangential morphology to the shell. 2) The filament may be located within the bubble, showing the general field morphology inside the \HII\ region. 3) The filament may be located inside or on the bubble, with parallel field lines formed by the filament's gas inflow (instead of by the \HII\ region).  In this case, the field lines may have transitioned from perpendicular to parallel (with respect to the filament) after gravity took over and material flowed along the filament~\citep[e.g.,][]{Liuetal2018, Busquet2020, Pillaietal2020}. Given that the overall magnetic field morphology of NGC~6334 appears to be perpendicular to the larger cloud, we suggest that the third scenario is most likely. 

\begin{figure}[hbpt]
\centering
\includegraphics[scale=0.43, trim={1.5cm 0.8cm 3cm 2.25cm}, clip]{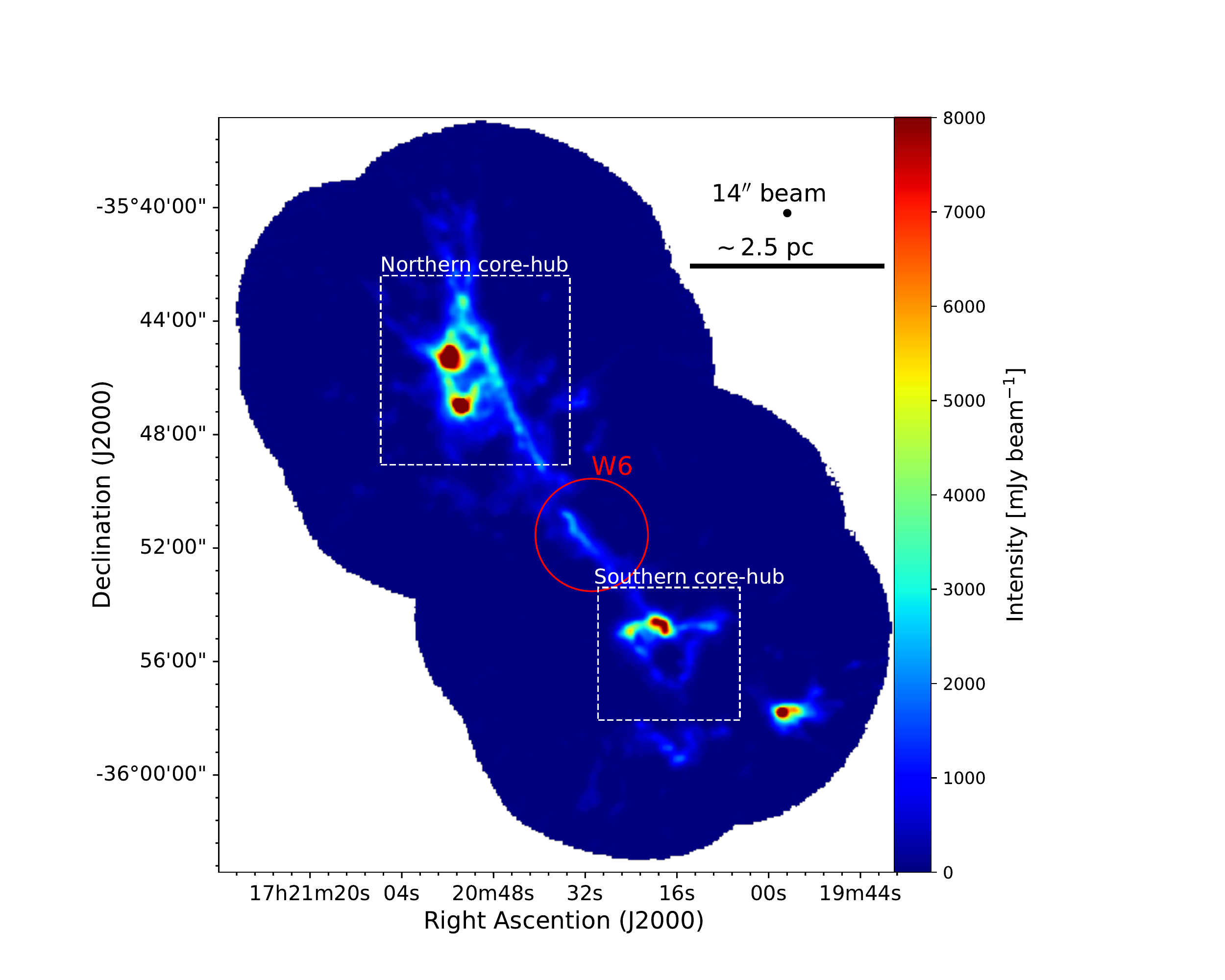}
\caption{Two dense star-forming regions within NGC~6334, each containing a number of cores. W6 is bounded on both sides by the two core-hub systems. }
\label{fig:NGC6334Cores}
\end{figure}

Additionally, given the presence of two dense regions (each containing numerous cores) to the north and south of this bubble (see Figure~\ref{fig:NGC6334Cores}), one could argue that the observed radial polarization pattern (tangential magnetic fields; see Figure~\ref{fig:RadialPol2column}) at the edge of this bubble is the result of material inflow to these two systems. We refer to these two regions as core-hub systems (see Figure~\ref{fig:NGC6334Cores}), and suggest that these tangential fields are caused by the interaction of the field lines with W6, not by the gravitational pull exerted by the two core-hub systems. This suggestion is backed up by the following arguments: First, we note that a magnetic field strength of  267\,$\mu$G is required to counteract the bubble's effects, while the average field strength of the filament is $\sim144\,\mu$G~(Paper~\RN{1}). Second, we estimate the gravitational pull exerted on W6 by the two dense core-hubs and find that it is one or two orders of magnitude lower than the gas or magnetic pressure of the bubble.

To calculate the gravitational pressure exerted  on the bubble by the hubs, we use
\begin{equation}
    P_{\rm{grav}} = \frac{GM\rho}{r}, 
\end{equation}
where $G$, $M$, $\rho$, and $r$ denote the gravitational constant, the mass of each of the hub systems (to the north or south), the volume density of the bubble, and the distance between the bubble and each core-hub center, respectively. We use the Virial  core masses (upper estimates) from \citet{Russeiletal2010} to approximate the mass of each core-hub region. 
The sums of all cores in the northern and southern hubs are 5270\,$M_{\odot}$ and 2124\,$M_{\odot}$, respectively, resulting in a  gravitational potential significantly less than the gas or the magnetic pressure. Even if we assume a mass of  10$^4$\,$M_{\odot}$ for each hub, a distance of  2\,pc from the center of the bubble, and a value of $\sim 10^3-10^4 \times m_p$ for $\rho$ (where $m_p$ is the mass of a proton), we find that gravitational pressure is one or two orders of magnitudes less than  gas or magnetic pressure. Therefore, we suggest that observed radial polarization in this region is caused by the bubble.

The center of \textbf{W7} (and the shell of S7) corresponds to the center of source~A, which has a magnetic field that varies across its diameter \citep[$\sim 200\arcsec$;][]{Sarmaetal2000}. The second strongest maser \citep[H$_2$O masers;][]{MoranRodiguez1980} in NGC~6334 and a Herbig-Haro-like object are both associated with this source, which has a bipolar structure, as illustrated in Figure~\ref{fig:RadialPol3column} (due to the \HII\ region being constrained by a dense molecular gas structure). The molecular toroid at the center of this source contains $\sim 2000\rm{M}_{\odot}$~\citep[][]{Sarmaetal2000} and may serve as a nursery for the formation of a protocluster of stars~\citep{Persietal2009}.  
Our results indicate that the observed Zeeman and DCF measurements are  smaller than the $B_{\rm{resist}}$ value that would be needed to provide support in this region by a factor of five (or an order of magnitude), thus the field lines are clearly altered by the two lobes, resulting in the observed radial polarization.

\textbf{W8} corresponds to the southernmost continuum source \citep[source~\RN{5};][]{McBreenetal1979}, which exhibits evidence for recent star formation activity \citep{KraemerandJackson1999} and is associated with H$_2$O and OH masers \citep{MoranRodiguez1980}.  This source contains a near-infrared bipolar nebula~\citep{Persi2019} and has no associated Zeeman magnetic field detections. 

To summarize, the  observations and analyses presented in this section strongly suggest that \HII\ regions are responsible for the radial polarization seen in W5 (and S5), W6, W7 (and S7-a), and W8. These bubbles show evidence of radial  polarization at a radius or a range of radii accompanied by increased polarization fraction, which is likely due to ordered magnetic fields. Additionally, some studies~\citep[][]{Hoangetal2019, Hoangetal2020, Trametal2021} predict an increase in polarization fraction as the distance from the center of the bubble increases, which can work together with ordered field lines and result in the observed locally maximal polarization fractions. 

\subsubsection{Energy balance of individual bubbles: Bubbles without radial polarization}

W2 (and S2-a), W3 (and S3) are located in a denser region of NGC~6334 and are spatially associated with sources E (S2-a) and F (S3). The magnetic field in the neighborhood of sources E and F exhibits an hourglass morphology~\citep{Zhangetal2014, Cortesetal2021}. Two possible explanations for the magnetic field morphology associated with these bubbles are as follows:  1) Because dust polarization observations are more sensitive to denser regions, we may be observing plane-of-sky magnetic fields associated with the foreground or background of these \HII\ regions, possibly accumulated on the shells; or 2) due to the higher densities in these regions, the ionized gas and outflow may lack the energy required to be dynamically important and alter the magnetic field morphology. Additionally, these two bubbles are younger and less evolved than other regions analyzed in this study. Therefore, they might not have had enough time to advance far enough into their cloud environment and reorder the field lines there.

\begin{figure*}[thbp]
\centering
\includegraphics[scale=0.4, trim={0cm 0cm 0cm 0cm},clip]{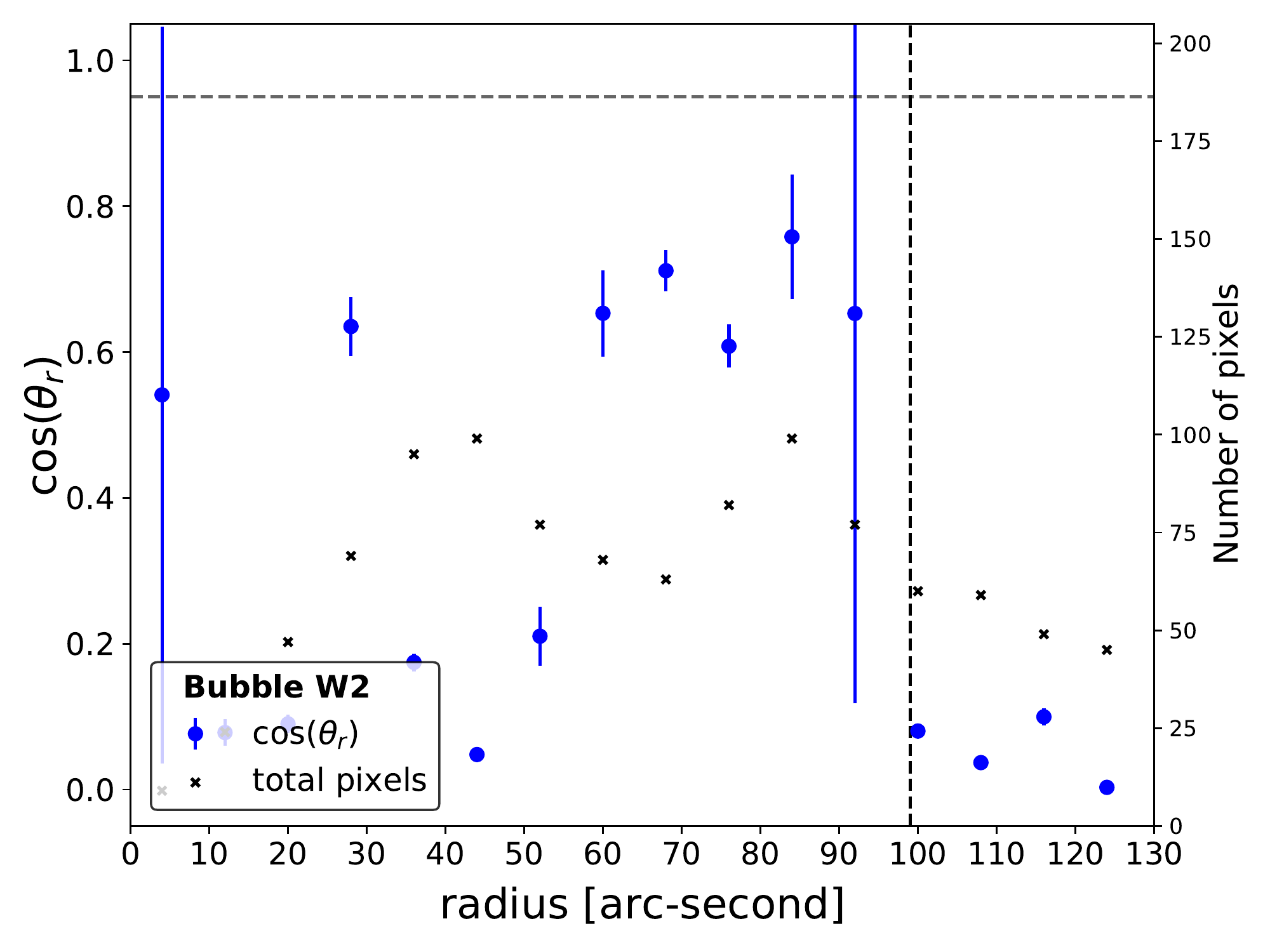}
\includegraphics[scale=0.4, trim={0cm 0cm 0cm 0cm},clip]{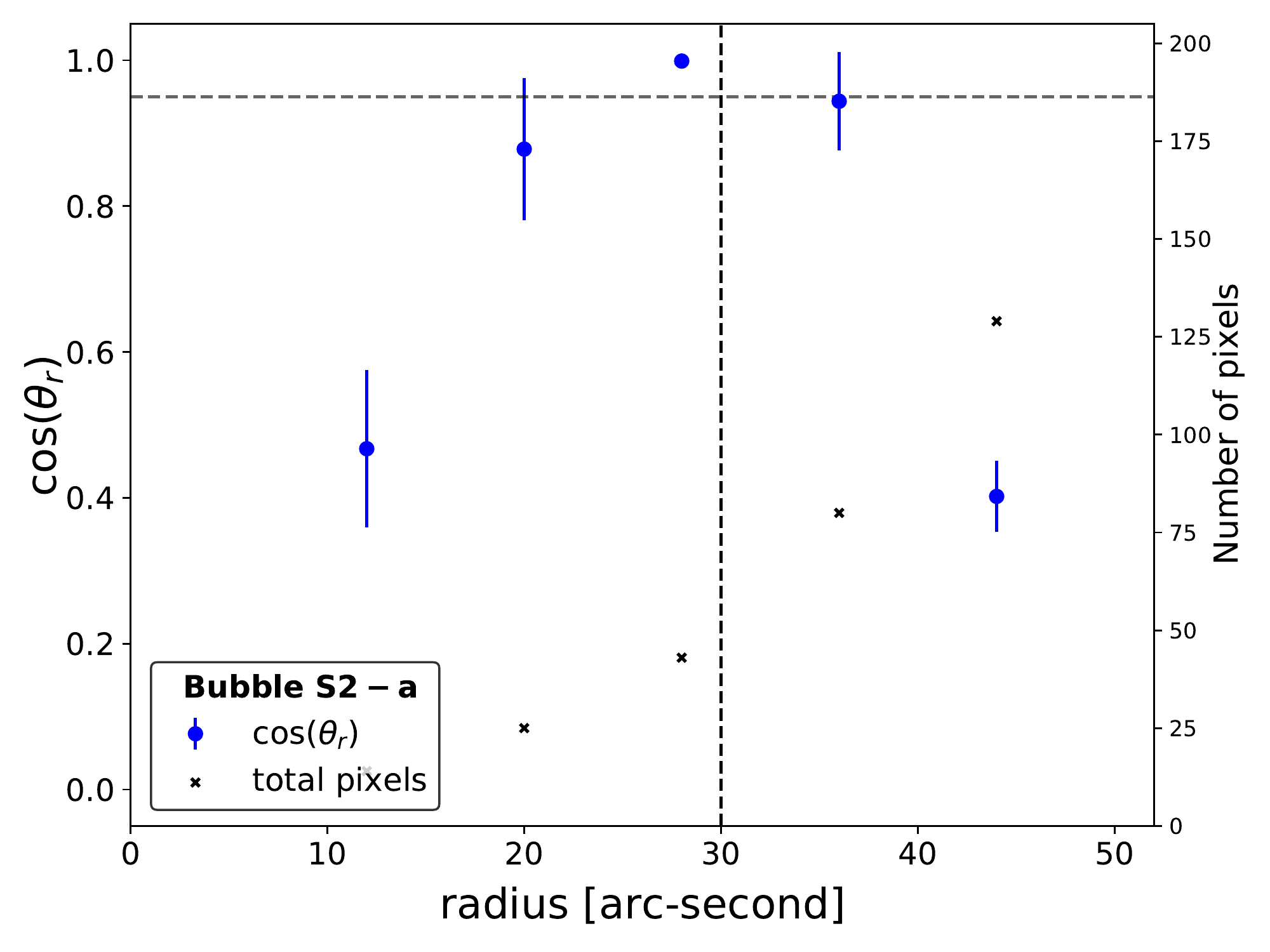}\\
\includegraphics[scale=0.4, trim={0cm 0cm 0cm 0cm},clip]{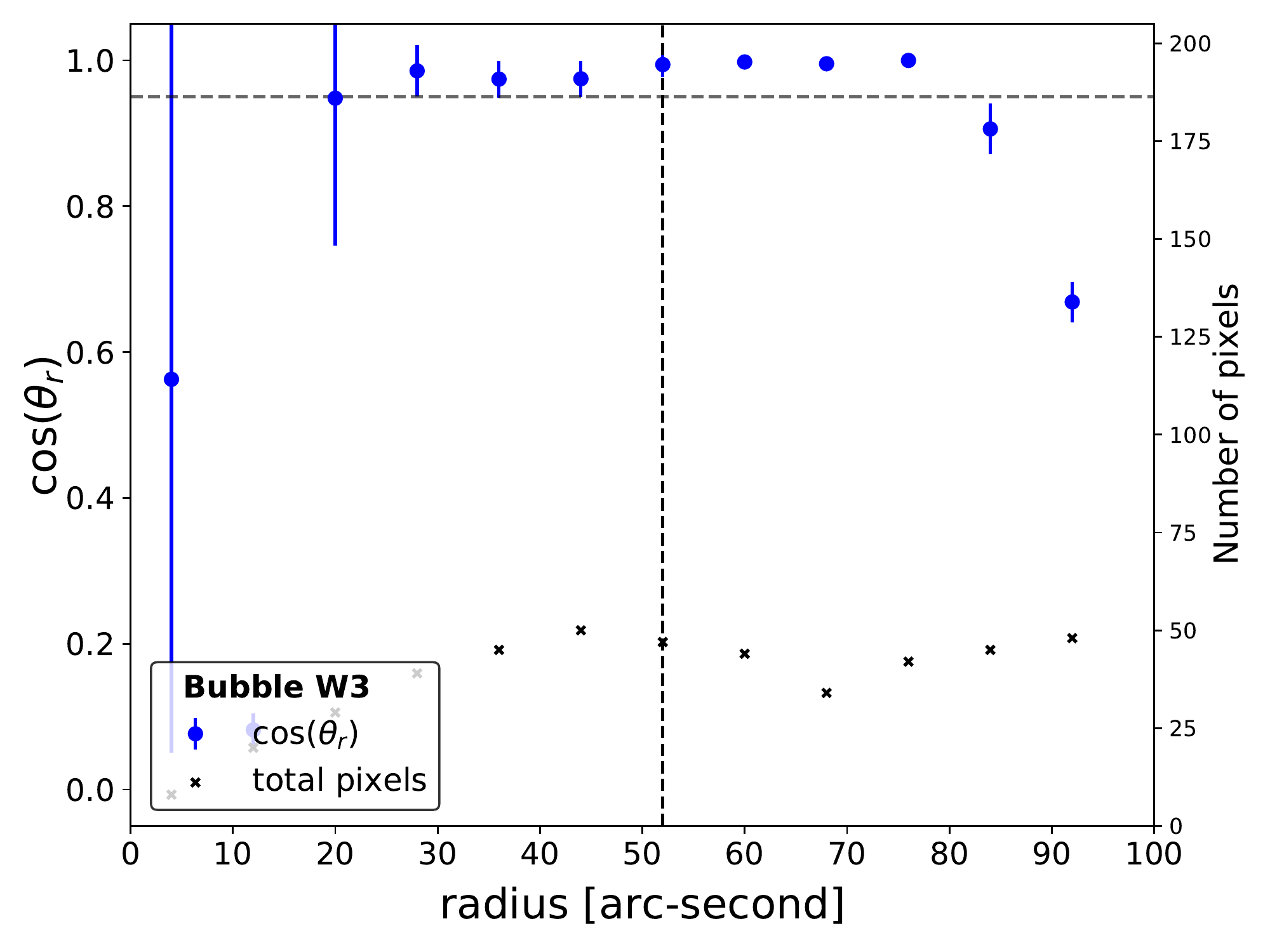}
\includegraphics[scale=0.4, trim={0cm 0cm 0cm 0cm},clip]{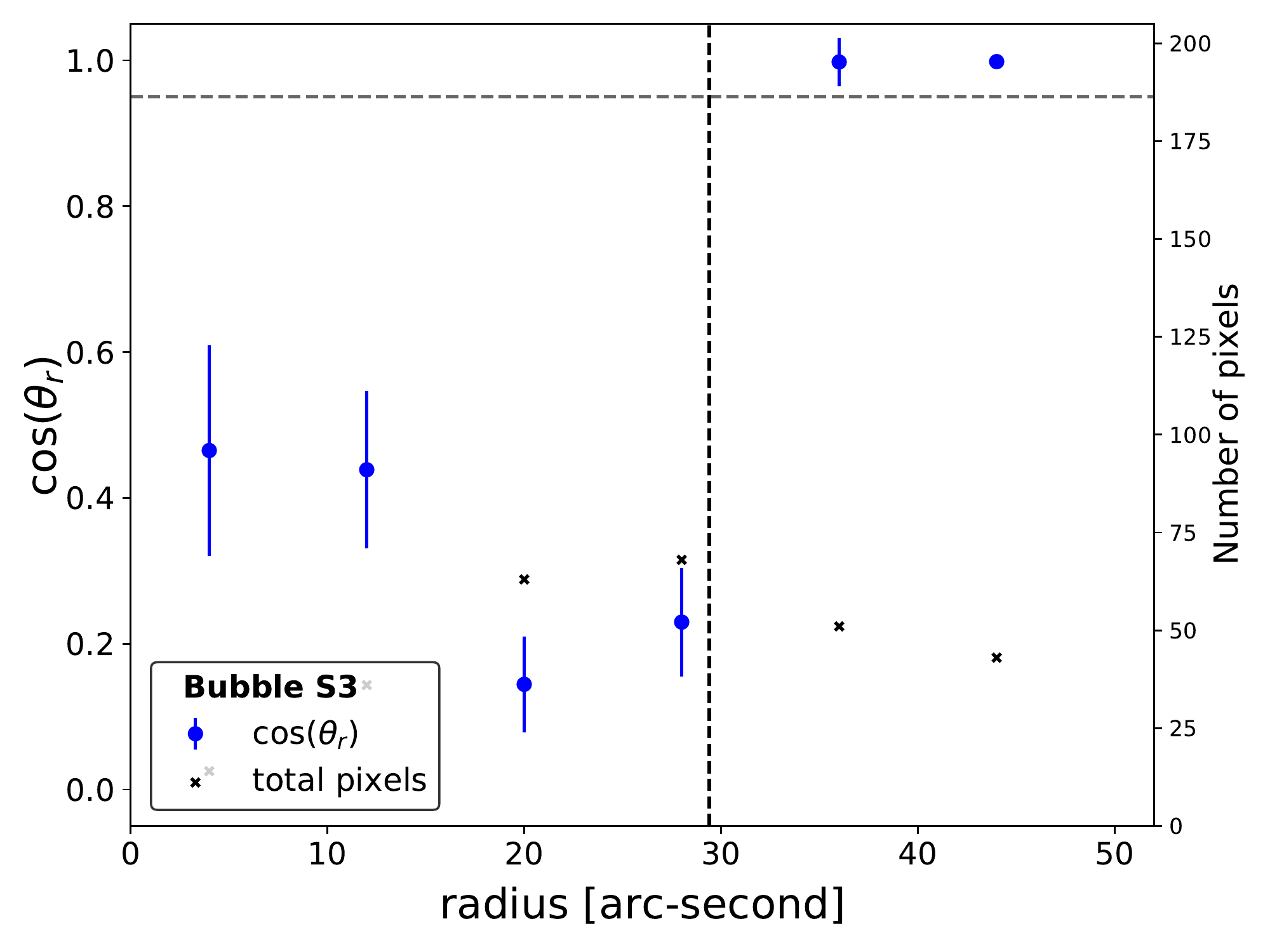}
\caption{Radial polarization analysis with  non-\HII\ effects subtracted for W2, S2-a, W3, S3. The vertical dashed line shows each bubble's radius identified by the WISE and Simpson catalogs. The black markers indicate the number of pixels contained in each annulus.  The blue points represent $\cos{\theta_r}$. In these plots, evidence of radial polarization is present at $27\arcsec$ for S2-a, between $\sim 33\arcsec$ to $45\arcsec$ for W3, and beyond $25\arcsec$ for S3. 
}
\label{fig:RadialPolBackSubt}
\end{figure*}

To explore these possibilities, we estimate the background and/or foreground (i.e., non-\HII ) contribution to the $I$, $Q$, and $U$ parameters in this neighborhood. To this end, we pick the interior region of S2-a that has a more coherent magnetic field morphology compared to the rest of the region (within $r \simeq 24\arcsec$) and average its Stokes $I$, $Q$, and $U$ values.  We then subtract these values from the corresponding observed values (in each pixel) and repeat the radial polarization analysis. The results are shown in Figure~\ref{fig:RadialPolBackSubt}, where W2 exhibits no sign of radial polarization again and W3 (between $30\arcsec$ to $75\arcsec$; and S3 outside the bubble) and S2-a (at 27\arcsec) show evidence of radial polarization.

The new (subtracted) radial polarization patterns associated with these bubbles may indicate that these bubbles have altered the field lines (first explanation). However, the observed magnetic field strengths are not small enough to confirm this. 
S2-a coincides with source E and its observed Zeeman measurements \citep[$340\pm 78$ with OH;][]{Sarmaetal2000} have a similar value as  $B_{\rm{resist}}$ (397\,$\mu$G). W3 and S3 correspond to source F, an ultra-compact \HII\ region with magnetic field strength of $\sim -2000$ to $-5000$\,$\mu$G~\citep[OH Zeeman;][]{Hunteretal2018}. 
This observed field value is greater (by a factor of two or more) than the region's  $B_{\rm{resist}}$ value of 744\,$\mu$G. These field values indicate that the bubbles had no observable impact on the field lines due to the high density of the region (second explanation). Higher resolution observations of these bubbles may shed light on the roles that these bubbles may have played in shaping the field lines in this region. 

\subsection{Machine vision algorithms to identify regions impacting field lines}
\label{sec:CHT}

Using published catalogs, we identified bubbles with radial polarization in Section~\ref{sec:radial}. In this section, 
we explore the feasibility of future studies using machine vision to identify regions with circular magnetic field morphologies (such as NGC~6334 \HII\ regions, where the field lines are oriented tangential to the bubble boundaries). 
To this end, we use the circular Hough transform technique (CHT).

CHT has been applied to a number of fields, most notably industrial applications such as eye detection~\citep[e.g.,][]{CHTEyeDetect} and traffic control~\citep[e.g.,][]{CHTTraffic}.  Edge detection algorithms are required when applying the CHT technique. To achieve the best results, we employ Canny edge detection\footnote{We use the already-existing edge-detection class in the Skit-image Python library~\citep{scikitimage}.}~\citep{Canny1986}, a numerical technique developed to optimize edge derivation. The Canny edge detection technique requires (as input) the smoothing width of the image as well as two threshold values for a double thresholding system (hysteresis thresholding), with the higher threshold typically set to be twice as high as the lower one.

Our approach includes the following two main steps, to locate regions that have pushed the magnetic field lines perpendicular to bubbles:
\begin{enumerate}
    \item We apply the CHT on two distinct maps: one is the presence of polarization data with SNR($I$)$ >10$ and SNR($PI$)$> 3$ (hereafter referred to as PolPres map), and the other is the polarization fraction (PF map; with the same selection criteria). The PolPres map has pixel values of 0 or 1 to indicate the presence or absence of data meeting the aforementioned selection criteria. The pixel values of the PF map are not binary and influence the edge detection. Different threshold values for both PolPres and PF maps, as well as the automatic Otso threshold determination\footnote{\url{skimage.filters.threshold_otsu}}, all yield results similar to the default threshold values of the  Canny edge detection library. 

    \item Using the circular Hough transform, we identify all possible circular patterns with radii ranging from $8\arcsec$ to $\sim 130\arcsec$, separately in each map. We then select those features that are identified in both maps at a similar location and radius (with a $\sim 40\arcsec$ difference allowance for their center location and a one-pixel difference allowance for their radius). Because some patterns are identified multiple times, we then average the radii and locations of bubble centers that are close together (within $20\arcsec$). We refer to the final selected circular features as CHT bubbles and denote them with the letter C followed by a number corresponding to the WISE bubbles.
\end{enumerate}

\begin{figure}[thbp]
    \centering
    \includegraphics[scale=0.47,clip, trim={1cm 0cm 2cm 1cm}]{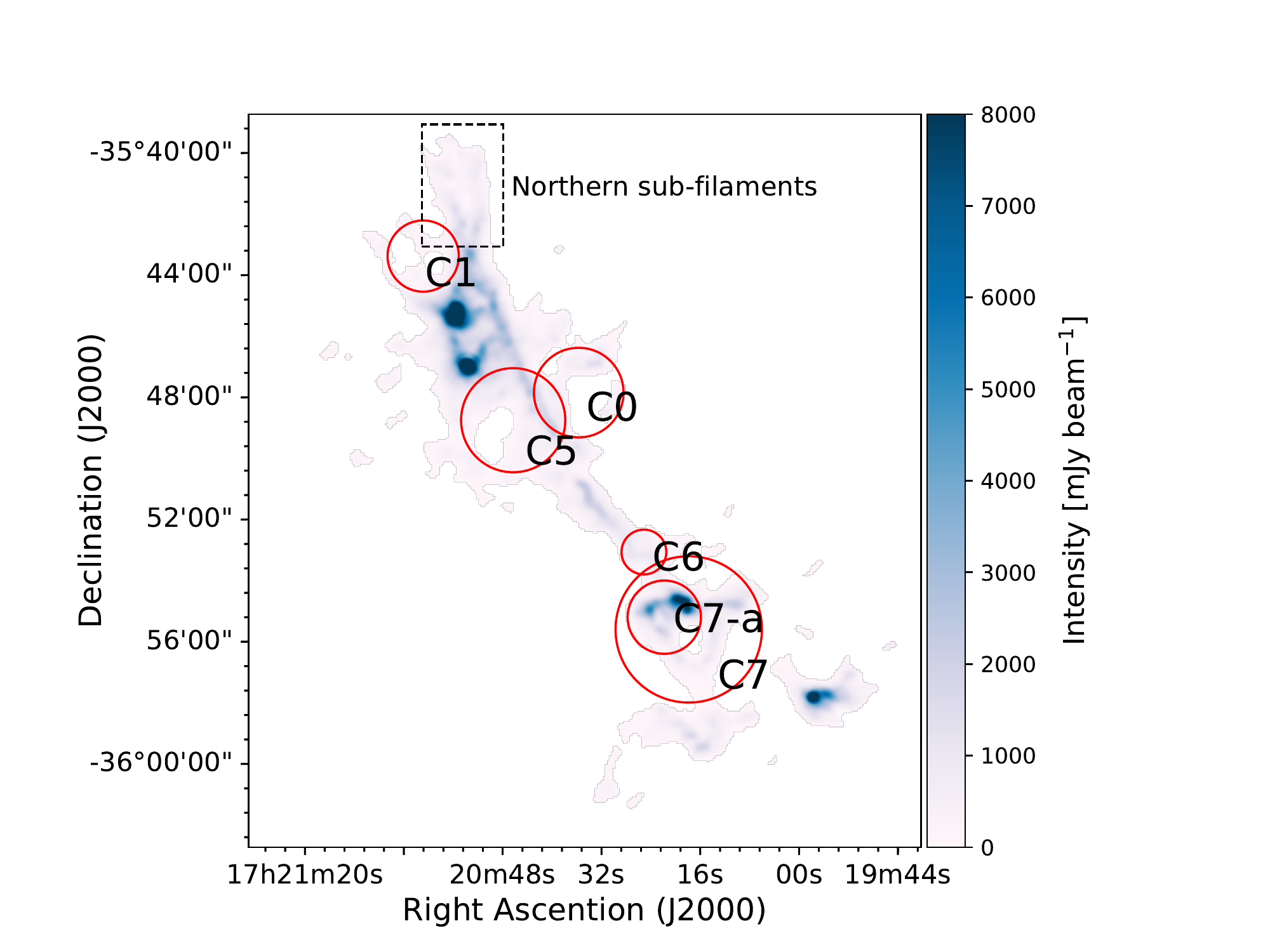}
    \caption{Identified bubbles using Circular Hough Transform. We identify bubbles that have altered the magnetic field lines. Some bubbles are identified multiple times, which are then averaged to one. The dashed black rectangle indicates the location of the two merging sub-filaments discussed in Paper~\RN{1}. }
    \label{fig:FinalCHT}
\end{figure}

In Section~\ref{sec:radial}, we found that W1, W5, W6, W7, and W8 exhibit radial polarization\footnote{W2, S2-a, W3, and S3 do not exhibit radial polarization in the observations.}, among which W1, W5, and W7 are the only ones with polarization data covering more than $\sim 20\%$ of the bubble's shell. Therefore, we propose that this technique should identify W1, W5, and W7.
As proposed, we find that the technique identifies W1, W5, and W7 (denoted as C1, C5, and C7, illustrated in Figure~\ref{fig:FinalCHT}). As a result, we believe the technique is promising and should be examined further with additional observations. 
Furthermore, the edges of W6 and S7-a (corresponding to C6 and C7-a) and an additional circular feature, C0, are identified in this technique. 
C6 coincides with the southern edge of W6 (a sub-region with clear radial polarization pattern), and C7-a overlaps with north-eastern side of S7-a. 

\begin{figure}[thbp]
\centering
\includegraphics[scale=0.4, trim={0cm 0.25cm 0cm 0.5cm},clip]{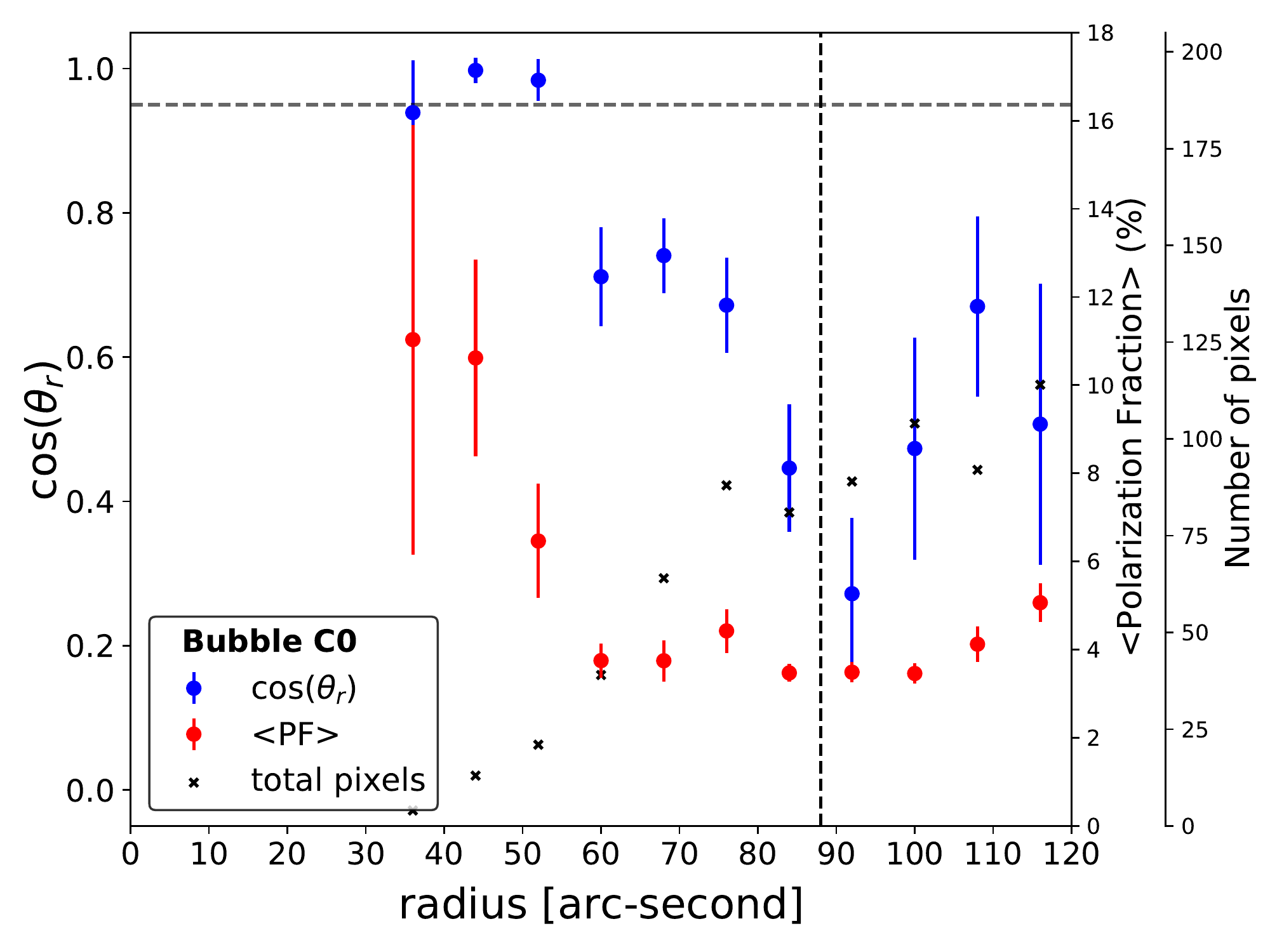}
\caption{Radial polarization analysis of C0. The mean polarization fraction, the total number of pixels in each shell (which satisfy the selection criteria of SNR($I$)$ >10$ and SNR($PI$)$> 3$), and $\cos(\theta_r)$ are all shown by the red, black, and blue markers.  }
\label{fig:C0}
\end{figure}

The magnetic fields associated with C0 exhibit a  radial polarization morphology between $\sim 45\arcsec$ and $53\arcsec$ (see Figure~\ref{fig:C0}). Because this region overlaps with W4, we suggest that C0 may be a distinct \HII\ region that was previously undetected due to line-of-sight confusion with W4. These findings indicate that while the CHT technique is used in a basic and simple form here, it has significant potential for detecting bubbles that have altered the magnetic field lines. These regions may be detected in full sky observations using more complex CHT techniques~\citep[e.g.,][]{ImprovedCHT} and/or machine learning.

\subsection{Cloud and magnetic field co-evolution}

\begin{figure}[thbp]
\centering
\includegraphics[scale=0.45,trim={0.cm 0.5cm 0.5cm 0.5cm}, clip]{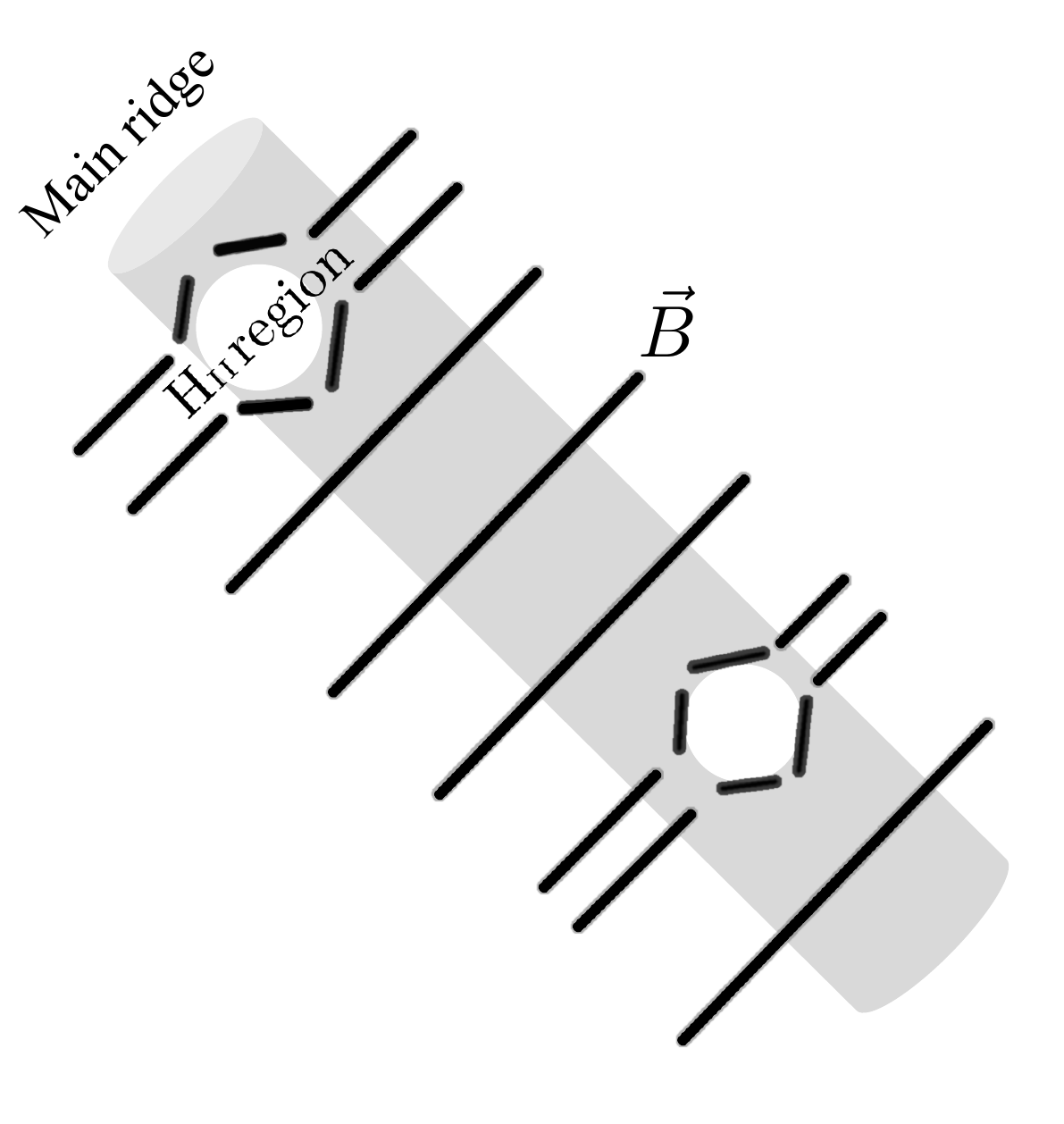}
\caption{A schematic illustration of perpendicular \bperp\ field lines on a dense molecular cloud with \HII\ region influences.  We propose that the plane-of-sky magnetic field lines in NGC~6334 are generally perpendicular to the main ridge and that the non-perpendicular lines are due to impacts by \HII\ regions or the gas flow (filament within W6). The cloud (or main ridge), magnetic field lines, and \HII\ regions are depicted with a cylinder, black lines, and circular cavities, respectively.}
\label{fig:conclusionCartoon}
\end{figure}

The large-scale \bperp\ lines are predominantly perpendicular to the main ridge of NGC~6334~\citep[][]{Lietal2015}, which are altered on substructure scales by \HII\ regions within the cloud. We suggest that these \bperp\ lines were initially even more coherent and perpendicular to the filament, but were then altered by feedback mechanisms, such as \HII\ regions, or gas flows (see the filament within W6), as illustrated in a schematic view by Figure~\ref{fig:conclusionCartoon}. 

Additionally, the physical properties of \HII\ regions may reveal the initial orientation of the magnetic field prior to the \HII\ expansion. For instance, studies~\citep[e.g.,][]{Chenetal2022} indicate that \HII\ regions may be elongated along the initial orientation of the field lines as expansion perpendicular to the field lines is met with resistance. \citet{Bonneetal2022} suggest that the initial magnetic field orientation may have also enforced the orientation of bipolar cavities in the Vela C cloud. We note that the elliptical elongations of the NGC~6334 bubbles are reported in the MIR observations of  \citet[][]{Simpsonetal2012}, but they are very small, except for S2-a as illustrated in Figure~\ref{fig:SimpsonEllipse} where the elongation of the bubble is not completely aligned with the large-scale magnetic fields~\citep[][]{Chenetal2022}.

The overall perpendicular \bperp\ lines to the main ridge (particularly at earlier evolutionary stages of the cloud) and the presence of large bubbles in this region, such as W5 may indicate that NGC~6334 formed via a cloud-cloud collision~\citep[][]{InoueFukui2013, Fukuietal2018PASJ, Fukuietal2021, Hayashi2021} or a shock-cloud interaction \citep[][]{Inutsukaetal2015, Inoueetal2018, Abeetal2021}. 
These cloud-formation models often result in an arc-shaped (bow-shaped) magnetic field morphology~\citep{Heiles1987, Tahanietal2019, Tahanietal2022O, Tahanietal2022P} that may appear perpendicular to the cloud when projected onto the plane of the sky. 
The shock-cloud interaction model is also supported in other regions by line-of-sight~\citep{Tahanietal2018, Tahanietal2020} and three-dimensional \citep{Tahanietal2019, Tahanietal2022O, Tahanietal2022P, Tahani2022} magnetic field and  velocity \citep{Arzoumanianetal2018, Bonneetal2020} observations.

\section{Summary and conclusion}
\label{sec:summary}
We investigated the magnetic field morphology of \HII\ regions associated with NGC~6334, using the JCMT BISTRO 850\,$\mu$m dust polarization observations. We located these bubbles using the WISE~\citep[][]{Andersonetal2014} and  \citet{Simpsonetal2012} catalogs and studied the presence of radial polarization patterns in each bubble by transforming the reference frame of the observations with respect to the bubble. This enabled us to identify bubbles that had altered their surrounding magnetic field morphology. 
Additionally, we compared the gas and magnetic pressures within each bubble using previously determined Zeeman magnetic field strengths in order to bolster our conclusion regarding the magnetic field morphologies associated with these bubbles.

We found evidence of radial polarization associated with W1, W5 (or S5), W6, W7 (or S7-a), and W8. The radial polarization is prominent in W5 (or S5) and W6, weak in W1 and S7-a, and absent in W2. S2-a and W3 (and S3) may indicate radial polarization after subtracting the non-\HII\ contributions.  
We propose that the magnetic field morphology of the NGC~6334 cloud was originally coherent and perpendicular to the cloud, but was altered during cloud evolution as a result of internal feedback mechanisms or gas flow.

We found that the presence or indication of radial polarization (tangential field lines) associated with \HII\ regions is accompanied by a higher polarization fraction. Quantifying the relationship between polarization fraction and bubble physical properties requires a statistical analysis of a large number of bubbles and should be investigated in subsequent studies. 

We found a bipolar polarization structure associated with W7, characterized by a clear radial polarization pattern for its northern lobe. To our knowledge, this is the first time that polarization data have been used to identify bubbles (or lobes within \HII\ regions). To take this a step further, 
we applied a computer vision approach based on the circular Hough transform on the dust polarization observations to identify and recognize bubbles that had altered their surrounding magnetic morphology. A more advanced computer vision or machine learning algorithm can enhance this technique, allowing it to be applied to additional regions or the entire sky.

\section*{acknowledgments}
This program is part of the JCMT BISTRO Large Program observed under project code M17BL011.
Team BISTRO-J is in part financially supported by 260 individuals. 
The James Clerk Maxwell Telescope is operated by the East Asian Observatory on behalf of the National Astronomical Observatory of Japan; Academia Sinica Institute of Astronomy and Astrophysics; the Korea Astronomy and Space Science Institute; Center for Astronomical Mega-Science (as well as the National Key R$\&$D Program of China with No. 2017YFA0402700). Additional funding support is provided by the Science and Technology Facilities Council of the United Kingdom and participating universities in the United Kingdom, Canada, and Ireland. Additional funds for the construction of SCUBA-2 and POL-2 were provided by the Canada Foundation for Innovation. 

The authors wish to recognize and acknowledge the very significant cultural role and reverence that the summit of Maunakea has always had within the indigenous Hawaiian community. We are most fortunate to have the opportunity to conduct observations from this mountain. 

We used \LaTeX, Python and its associated libraries including astropy~\citep{astropy}, PyCharm, Jupyter notebook, SAO Image DS9, and the Starlink~\citep{Starlink} software.  QuillBot\footnote{\url{https://quillbot.com/}} was employed to edit the text. 

M.Tahani is supported by the Banting Fellowship (Natural Sciences and Engineering Research Council Canada) hosted at Stanford University and the Kavli Institute for Particle Astrophysics and Cosmology (KIPAC) Fellowship. D.J. is supported by the National Research Council of Canada and by a Natural Sciences and Engineering Research Council  Discovery Grant. T.O. acknowledges a Grant-in-Aid for Scientific Research of JSPS (18K03691). Mo.Ta. is supported by JSPS KAKENHI grant Nos.18H05442, 15H02063, and 22000005. J.K. is supported by JSPS KAKENHI grant No.19K14775. W.K. is supported by the National Research Foundation of Korea (NRF) grant funded by the Korea government (MSIT) (NRF-2021R1F1A1061794). F.P. acknowledges support from the Spanish State Research Agency (AEI) under grant number PID2019-105552RB-C43. C.E. acknowledges the financial support from grant RJF/2020/000071 as a part of the Ramanujan Fellowship awarded by the Science and Engineering Research Board (SERB), Department of Science and Technology (DST), Government of India. L.F. acknowledges support from the Ministry of Science and Technology of Taiwan, under grants no. MoST107-2119-M-001-031- MY3, 111-2811-M-005-007 and 109-2112-M-005-003-MY3, and from Academia Sinica under grant no. AS-IA-106-M03. F.K is supported by the Spanish program Unidad de Excelencia María de Maeztu CEX2020-001058-M, financed by MCIN/AEI/10.13039/501100011033.

\bibliography{bibt}{}
\bibliographystyle{aasjournal}

\appendix
\section{Error bars}
\label{apndx:radialPolErrorBar}

To determine the error bars associated with $\cos(\theta_r)$ we use the following error propagation equation: 
\begin{equation}
    \delta f = \sqrt{(\frac{\partial f}{\partial Q_r})^2 (\delta Q_r)^2 + (\frac{\partial f}{\partial U_r})^2 (\delta U_r)^2},
\end{equation}
where $\delta Q_r$ and $\delta U_r$ represent the standard deviation of the mean of $<Q_r>$ and $<U_r>$ of each annulus, respectively, and
\begin{equation}
\begin{aligned}
\centering
    f &= \cos{\theta_r}, \\
    \theta_r &= 0.5 \times \arctan(<U_r>, <Q_r>), \\
    \frac{\partial f}{\partial U_r} &= -0.5 \times \sin{\theta_r} \times \frac{1}{<Q_r>\big(1+(\frac{<U_r>}{<Q_r>})^2\big)}, \\
    \frac{\partial f}{\partial Q_r} & = 0.5 \times \sin{\theta_r} \times \frac{1}{1+(\frac{<U_r>}{<Q_r>})^2}, \times \frac{<U_r>}{<Q_r>^2}. 
\end{aligned}
\end{equation}

\section{Elliptical bubble}

\begin{figure}[thbp]
\centering
\includegraphics[scale=0.5,clip, trim={1cm 0cm 2.5cm 1.5cm}]{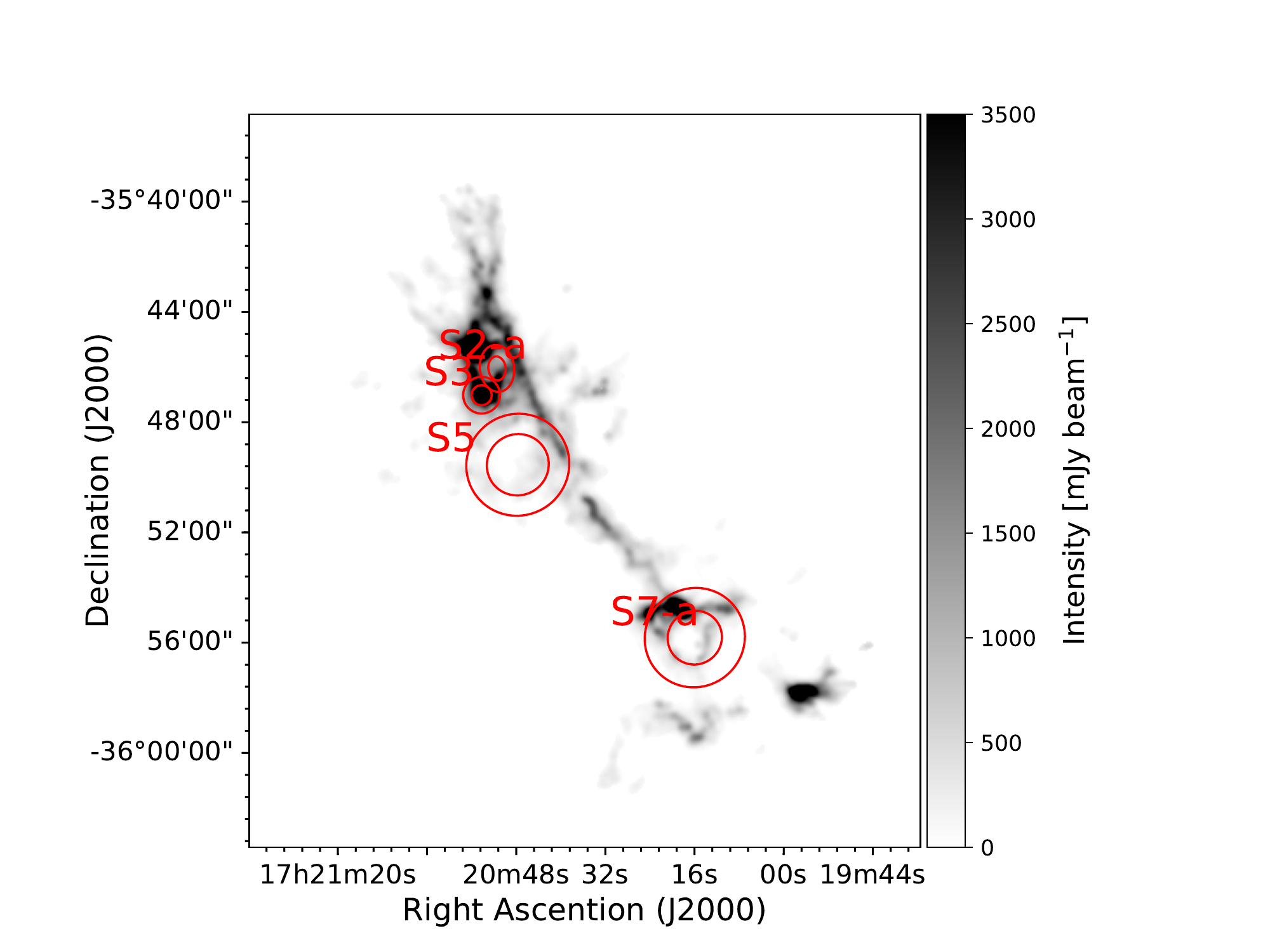}
\caption{The Simpson bubbles illustrated in their elliptical form. The numbers show the Simpson bubbles specified in Figure~\ref{fig:NGC6334Bubbles}. The inner and outer radii corresponding to each bubble are illustrated with red ellipses.}
\label{fig:SimpsonEllipse}
\end{figure}

The bubbles in the \citet[][]{Simpsonetal2012} catalog are initially classified as ellipses. The Simpson catalog includes inner and outer radii for both X and Y directions, as well as their eccentricity values, as illustrated in Figure~\ref{fig:SimpsonEllipse}. We examined the radial polarization patterns of these regions and found that they were very similar to the results presented in Figures~\ref{fig:RadialPol3column} and \ref{fig:NoRadialPol}, due to the small eccentricity values of these bubbles. 

\end{document}